\documentclass[a4paper,english,onecolumn, 12pt, draftclsnofoot]{IEEEtran}
\usepackage[T1]{fontenc}
\usepackage[latin9]{inputenc}
\usepackage{amsmath}
\usepackage{amssymb}
\usepackage{graphicx}
\usepackage{esint}

\makeatletter

\pdfpageheight\paperheight
\pdfpagewidth\paperwidth

\providecommand{\tabularnewline}{\\}

\usepackage[T1]{fontenc}
\usepackage[latin9]{inputenc}
\usepackage{comment}
\usepackage{amsmath}
\usepackage{float}
\usepackage{amssymb}
\usepackage{graphicx}
\usepackage{enumitem}
\usepackage{physics}
\usepackage{cite}

\usepackage{pstricks}
\usepackage{indentfirst}
\usepackage{amsthm}
\usepackage{glossaries-extra}
\setabbreviationstyle[acronym]{long-short}
\newacronym{CRAN}{CRAN}{Centralized radio access network}
\newacronym{RUs}{RUs}{remote units}
\newacronym{RU}{RU}{remote unit}
\newacronym{mmW}{mmW}{millimeter-wave}
\newacronym{PPP}{PPP}{Poisson point process}
\newacronym{SG}{SG}{Stochastic Geometry}
\newacronym{CF}{CF}{characteristic function}
\newacronym{MRC}{MRC}{Maximal Ratio Combining}
\newacronym{SC}{SC}{Selection Combining}
\newacronym{PGFL}{PGFL}{probability generating functional}
\newacronym{LOS}{LOS}{line-of-sight}
\newacronym{SE}{SE}{spectral efficiency}
\newacronym{BS}{BS}{base station}
\newacronym{AP}{AP}{access point}
\newacronym{NLOS}{NLOS}{non-line-of-sight}
\newacronym{SINR}{SINR}{signal-to-noise-and-interference}
\newacronym{PDF}{PDF}{probability density function}
\newacronym{mMIMO}{mMIMO}{massive multiple-input multiple-output}
\newacronym{CDF}{CDF}{cumulative density function}
\newacronym{UcylA}{UcylA}{uniform cylindrical array}
\newacronym{UPA}{UPA}{uniform planar array}
\newacronym{UCA}{UCA}{uniform circular array}
\newacronym{EIRP}{EIRP}{equivalent isotropically radiated power }
\newacronym{ULA}{ULA}{uniform linear array}
\newacronym{UEs}{UEs}{user equipments}
\newacronym{BF}{BF}{beamforming}
\newacronym{UE}{UE}{user equipment}
\newacronym{UAV}{UAV}{unmanned aerial vehicles}
\newacronym{MIMO}{MIMO}{multiple-input multiple-output}
\newacronym{CU}{CU}{central unit}
\newacronym{BW}{BW}{bandwidth}
\newacronym{DAS}{DAS}{distributed antenna systems}
\newacronym{FSPL}{FSPL}{free-space path-loss}
\newacronym{SIR}{SIR}{signal-to-interference ratio}

\usepackage[framemethod=TikZ]{mdframed}
\usepackage{lipsum}
\mdfdefinestyle{MyFrame}{%
    linecolor=blue,
    outerlinewidth=2pt,
    roundcorner=20pt,
    innertopmargin=\baselineskip,
    innerbottommargin=\baselineskip,
    innerrightmargin=20pt,
    innerleftmargin=20pt,
    backgroundcolor=white!50!white}

\makeatother

\usepackage{babel}

\newcommand{\thicktilde}[1]{\mathbf{\tilde{\text{$#1$}}}}

\begin{document}

\title{{\Large{Performance of dense wireless networks in 5G and beyond using stochastic geometry}}}
\author{\emph{Reza Aghazadeh({*}) and Umberto Spagnolini}\\
\emph{ \{Reza.Aghazadeh@polimi.it;Umberto.Spagnolini@polimi.it\}}}

\maketitle
Dep. Elettronica, Informazione and Bioingegneria, Politecnico di Milano,
Milan (Italy)
({*}) corresponding author

\begin{abstract}
Device density in cellular networks is expected to increase considerably in the next future. Accordingly, the access point (AP) will equip  massive multiple-input multiple-output (mMIMO) antennas, using collimated millimeter-wave (mmW) and sub-THz communications, and increase the bandwidth to accommodate the growing data rate demands. In this scenario, interference plays a critical role and, if not characterized and mitigated properly, might limit the performances of the network. In this context, this paper derives the statistical properties of the aggregated interference power for a cellular network equipping a mMIMO cylindrical array. The proposed statistical model considers the link blockage and other network parameters such as antenna configuration and device density. The findings show that the characteristic function (CF) of the aggregated interference power can be regarded as a weighted mixture of two alpha-stable distributions. Furthermore, by analyzing the service probability, it is found that there is an optimal configuration of the array depending on the AP height and device density. The proposed statistical model can be part of the design of dense networks providing valuable insights for optimal network deployment.
\end{abstract}
\begin{center}
\textbf{Keywords:} 5G- mmW, 6G systems, Poisson point process, Interference characterization, Stochastic geometry, Outage
analysis, Beamforming, Uplink, Uniform cylindrical array, Blockage
\par\end{center}
\section{Introduction}

The fifth-generation (5G) cellular network has been recently deployed with unprecedented communication performance, i.e., 10-100X times higher data rate, 1ms latency, and much higher area throughput \cite{PANWAR201664,Rappaport2013MillimeterWM}. The upcoming sixth-generation (6G) cellular network promises to further improve current performance by at least one order of magnitude \cite{6GSpectrum}. To meet such requirements, it is necessary to operate on multiple frontiers, e.g., increase the bandwidth, cell density, transceiver efficiency.

Current cellular networks operate at sub-6GHz band, which is heavily congested \cite{6GSpectrum}. Recently millimeter waves (mmW) and sub-THz frequencies ($30 - 300$GHz) have gained a substantial interest because of the large unexploited spectrum \cite{5GSpectrum2,5GSpectrum3,THzArticle}. However, propagation at these frequencies experiences higher path and penetration loss making the links prone to blockage \cite{BlockageKai}. 
A solution to these challenges is to use beam-type communication based on massive \gls{mMIMO} systems and increase cell density \cite{SpectralEfficiency}. However, as the device density increases, interference emerges as one of the main challenges to be characterized and mitigated. 
Characterization of the aggregate interference power and coverage
analysis in \gls{mmW} networks has been investigated over the past
years but only in some specialized settings, under the \gls{LOS} and \gls{NLOS} propagation. Win et. al. \cite{WinPintoMatThInterference}
derived the distribution of the aggregate interference power and amplitude for PPP distributed \gls{UEs} on a 2D plane when the antenna is omnidirectional and co-planar with \gls{UEs} (array's height $h=0$). The impact analysis 
of the users' height has been studied in ref.\cite{UsersHeightEffectSG}.
The array's height $h$ plays an important role that needs to be analyzed, namely in view of the 5G \gls{mmW} and 6G use-cases. Therefore, it is necessary to extend the \gls{SG} framework to the 3D framework of antennas and \gls{UEs} (see e.g. \cite{DenseWN3dSG}). \gls{SG} provides a preferred framework in network modeling to perform coverage and rate performance analysis \cite{WinElSawyTutorial1,SGHaeneggiTutorial,DBLP:Andrews,HeathUplinkSINRRateSG,SGDiRenzo}. The impact of the antennas' height in a 3D SG for ultra-dense networks proves that there is an upper limit on network performance which is dependent on the path-loss model parameters \cite{AntennaHeightImpact}. Even if the impact of the antenna and \gls{UE} height difference has been studied \cite{liu2017impact}, the existence of an optimum array height has not been deduced in dense networks. Here we derive the analytical model of aggregated interference and show that the optimum array height depends on the path-loss model but also on the users' density, array type, and size. Similarly, the impact of the height in low-altitude aerial platforms
\cite{Low-AltitudeAerialplatforms(LAPs),3DUAVOptimumHeight} and in unmanned aerial vehicles \cite{CoexistenceTerrestrialAerial,3DUAV} proves that there are optimum altitudes maximizing the coverage probability according to some specific scenarios.
\begin{figure}[tbh!]
\begin{centering}
\includegraphics[scale=0.7]{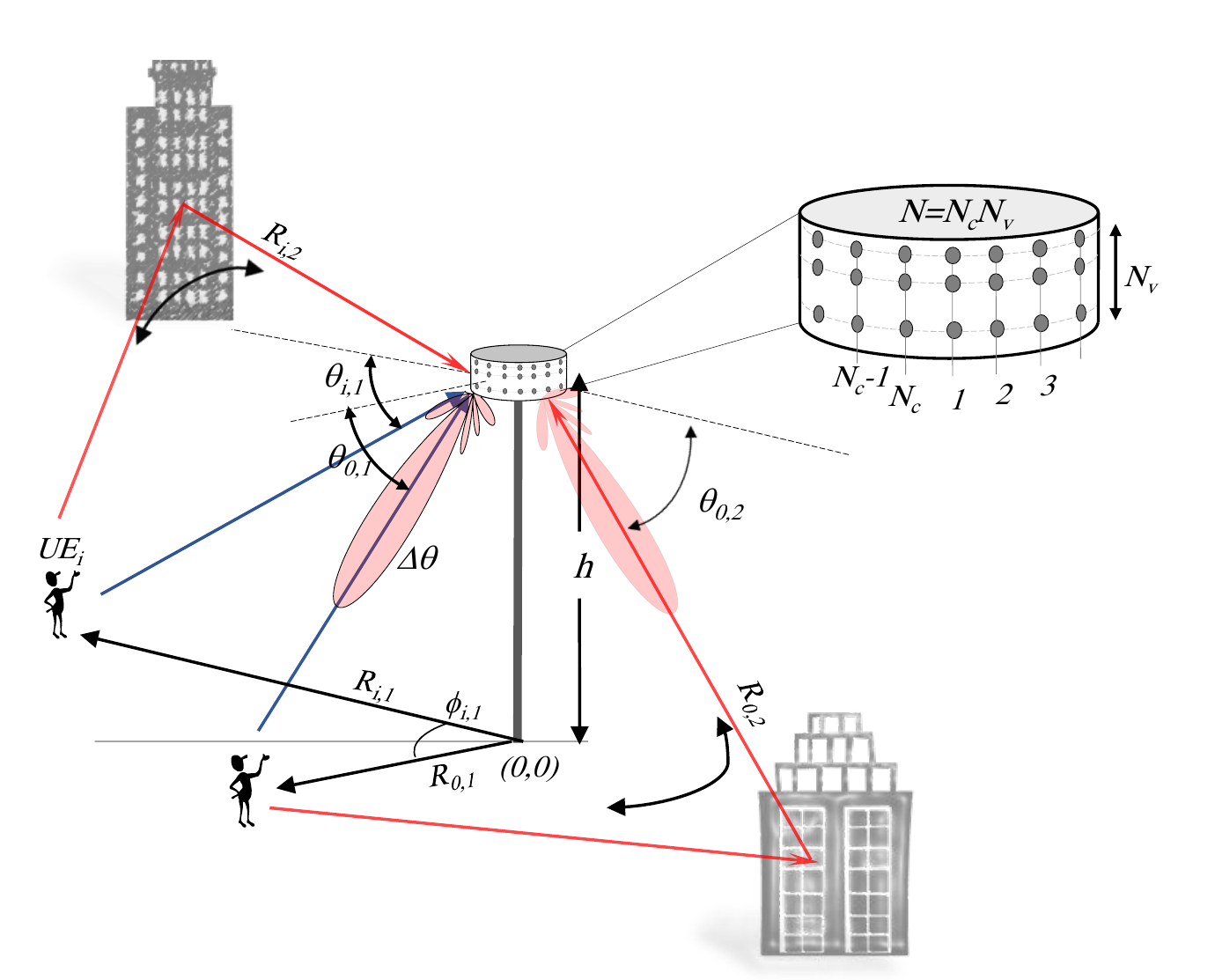}
\par\end{centering}
\caption{\label{fig:Cylindrical-array-configuration}Configuration of a $N_{c}\times N_{v}$ uniform cylindrical array (UcylA) with line-of-sight (LOS) and non-line-of-sight (NLOS) links: ($\phi_{i,\ell}$,$\theta_{i,\ell}$) are the azimuth and elevation angels, $h$ is the height of the array and the pointing directions are toward all the LOS ($\ell = 1$) and NLOS ($\ell = 2,3,\cdots$) links arriving from the user equipment (UE) over a propagation oath $R_{i,\ell}$.}
\end{figure}
In the literature, it is common to assume that the fading follows Rayleigh distribution (or at least the interference link has a Rayleigh
distribution) or Nakagami-m distribution. The coverage probability
has a tractable form as a Laplace function of the aggregate interference
power (see, e.g. \cite{WinElSawyTutorial1,DianaDanielaCoverageCooperation,CoverageImperfectAlignmentChinese,AndrewsBacelliTractableApproach}).
However, for general fading, using the Laplace function of
the aggregate interference is not possible anymore and thus, the coverage
and rate analysis cannot be expressed in a tractable way. To overcome
this problem, ref. \cite{SGHaeneggiTutorial} introduces multiple
techniques, including some methods to find the \gls{PDF} of the aggregate
interference calculated from the corresponding \gls{CF}.

In previous works, interference distribution for single antennas or
the \gls{ULA} (e.g. in \cite{DianaDanielaCoverageCooperation}), configuration were investigated most. The contribution
of this paper is the usage of the 3D \gls{SG} framework in the uplink for homogeneous \gls{PPP} with a density $\lambda$ on a pseudo-3D geometry  (sometime referred to as 2.5D geometry \cite{25D_1}, \cite{25D_2}) where \gls{UEs} lie in a plane (say ground) and the \gls{AP} with $N_{c} \times N_{v}$ \gls{UcylA} (i.e. a set of $N_{v}$ half-wavelength rings of $N_{c}$ \gls{UCA} antennas/each) has the height $h\geq0$. The Fig. \ref{fig:Cylindrical-array-configuration} illustrates the setup for the computation of the properties of the aggregate interference from the ensemble of $UE_1,UE_2,...UE_i,...$ when the array at the \gls{AP} is pointing toward the user of interest $UE_0$. We show that being able to characterize the aggregated interference, one can optimally adjust the configurations of the \gls{AP} to increase the number of users served and accordingly the coverage probability.  

We analytically prove in Sect.\ref{sec:InterfCharact} that the aggregated interference power onto an array of antennas located in an arbitrary height $(h)$ has a \gls{CF} that can be decomposed into  a mixture of two stable distributions (a skewed alpha-stable and a Gaussian distribution). The closed-form \gls{CF} for $h>0$ is another novelty to compute the \gls{PDF} or \gls{CDF} of the aggregated interference power, without cumbersome numerical integration. The scenario considered herein for users and interferers accounts either the LOS and NLOS propagation, and also the possibility of link blockage as typical in mmW and sub-THz systems. The analytical model enables (Sect.\ref{sec:Numerical-results}) the evaluation of the trade-off in arrays' height selection, and the impact of the propagation scenarios.
The \gls{UcylA} with $h>0$ generalizes the previous works on \gls{ULA} configuration that was investigated mostly for coverage analysis (see, e.g.,
\cite{DianaDanielaCoverageCooperation,DirectionalArrayImpactHaenggi}). On array engineering, \gls{UcylA} can be designed by multifaceted array as far more practical, and results in  \cite{josefsson2006conformal} supports the conclusion that any results for curved arrays apply to faceted ones.

Based on the analytical model, the main results can be summarized as follow: i) in most  scenarios with small interferers' density it is more beneficial to adopt a \gls{UCA} rather than \gls{UcylA} while for large interferers' density a \gls{UcylA} would be preferred; ii) there is an optimum \gls{AP} height that depends on propagation and interferers' density $\lambda$; iii) at \gls{AP} height $h=0$ the aggregated interference power is alpha-stable distributed, and for $h \rightarrow \infty$ the limit becomes Gaussian, but for any arbitrary height $h$ it is decomposable into two stable distributions; iv) blockage probability is impacting the service probability for small $\lambda$, but less if counting the average number of users served within a region;
v) a connection with multiple paths (LOS and NLOS) is more beneficial for small $\lambda$, as for  large $\lambda$ the interference is too large.

The paper is organized as follows. We present the system model including the signal and array gain models in Sect.\ref{sec:System-model}. The \gls{CF} of the aggregate interference for \gls{UcylA} in \gls{LOS} links in Sect.\ref{sec:CF}. 
Sect \ref{sec:InterfCharact}  contains the statistical characterization of the interference power. In Sect.\ref{sec:NLOS} the \gls{CF} is extended considering NLOS paths, noise power and blockage, and we conclude the paper in Sect.\ref{sec:Conclusion}.

\section{System model\label{sec:System-model}}

The scenario is in Fig. \ref{fig:Cylindrical-array-configuration} where the \gls{AP} has
a cylindrical array with $N=N_{c}\times N_{v}$ antennas in total. The \gls{UEs}
are uniformly distributed following a homogeneous \gls{PPP} with density $\lambda$ that denotes the mean number of active \gls{UEs} per
square meter.
The spatial channel of the \gls{mmW} and 6G sub-THz systems are purely directional (see, e.g., \cite{ModelingMillimeterWaveCellularSystems}),
and the \gls{LOS} (or \gls{NLOS}) link is  affected by the path-loss modeled in terms of UE-AP distance $d_{LOS}$ (or $d_{NLOS}$), and faded amplitude $\beta_{LOS}$ (or $\beta_{NLOS}$). The propagation attenuation model for \gls{LOS} and \gls{NLOS} is $\beta/d^{b}$ with amplitude path-loss $b\geq1$. 
The array of antennas is uniformly cylindrical, the isotropic radiating antennas are arranged into a set of  $N_{v}$ \gls{UCA}s with $N_{c}$ antennas each and antennas' spacing is half the wavelength. Namely, the two arrangements of antennas are such that the corresponding beamforming of \gls{UCA}s reduce the interference angularly, and the vertical arrangement of the rings (acting as vertical \gls{ULA}s), tilts the beam to improve the capability to reduce the near interferers when pointing toward far-away \gls{UEs}. Note that here the UEs are considered on the ground, which means that the \gls{AP} height is the height difference of UEs and \gls{AP} (pseudo 3D or 2.5D geometry).

\subsection{Array gain model\label{subsec:Array-gain-model}}
Each \gls{AP} equipped with the array of antennas is positioned at height
$h$ from the ground at $(0,0)$ planar coordinates as in Fig. \ref{fig:Cylindrical-array-configuration}. The array gain
for the \gls{UcylA} in far-field $G(\phi,\theta)$
depends on elevation angle ($\theta$) and azimuth ($\phi$),
which in turn depends on the number of antennas partitioning between
$N_v$ and $N_c$. The beamforming for the cylindrical array is
conveniently decomposed into the design of two compound arrays, and
thus the array gain $G(\phi,\theta)=G_{c}(\phi)G_{v}(\theta)$
is separable into the \gls{UCA} gain $G_{c}(\phi)$ and vertical \gls{ULA}
gain $G_{v}(\theta)$  \cite{685696,3DBF}. The approximation
holds true in \gls{UcylA} when using separable weightings \cite{SeparationOfWeightingUcylA,VanTrees2001}. The beamforming used here is the conventional one that is optimum
for uniform interference, and the array gains for half-wavelength
inter-element spacing either for \gls{UCA} and \gls{ULA} are \cite{VanTrees2001,spagnolini2018statistical}:
\begin{eqnarray}
G_{c}(\phi) & = & J_{0}\left(\frac{N_{c}}{2}\sqrt{(\cos\phi-\cos\phi_{o})^{2}+(\sin\phi-\sin\phi_{o})^{2}}\right)\label{eq:directivity-patternUCA}\\
G_{v}(\theta) & = & \frac{\sin[\pi(\sin\theta-\sin\theta_{o})N_{v}/2]}{N_{v}\sin[\pi(\sin\theta-\sin\theta_{o})/2]}\label{eq:directivity-patternULA}
\end{eqnarray}
where ($\phi_{o},\theta_{o}$) denotes the pointing azimuth
and elevation pair to the intended \gls{UE}, and for \gls{UCA} the gain approximation \ref{eq:directivity-patternUCA} is by
the zero-order Bessel function of the first kind $J_{0}(.)$ that
can be shown to be accurate for $N_{c}\geq16$. Note that array gains
are normalized for convenience to have $G(\phi_{o},\theta_{o})=1$. The beam width along the two angles, $\Delta\phi$, and $\Delta\theta$, are inversely proportional to $N_{c}$ and $N_{v}$, respectively. Elevation beam width is further distorted by the effective array aperture that makes the beam width scale with the cosine of the tilt angle (stretching effect): $\Delta\theta/\cos(\theta-\theta_{o})$.

\subsection{Signal model and service probability}
Let $x$ be the transmitted signal, the signal received by the
\gls{AP} with beamforming pointing toward the \gls{UE} of interest
with angles $\phi_{o}=0$ and radial distance $R_o$ is 
\begin{equation}
y=\frac{\beta_{o}}{\left({R_{o}^{2}+h^{2}}\right)^{\frac{b}{2}}}G(\phi_0,\theta_0)x+ \iota + w \label{eq:SimpleSignalModel},
\end{equation}
where ${w}\sim CN(0,\sigma_{w}^{2}/N)$ is the additive Gaussian
noise with power $\sigma_{w}^{2}/N$ after the array gain, and 
\begin{equation}
    \iota = \sum_{i=1}^{\infty}\frac{\beta_{i}}{\left({R_{i}^{2}+h^{2}}\right)^{\frac{b}{2}}}G(\phi_{i},\theta_{i})x_{i}\label{eq:LOSInterfAmplitude}.
\end{equation}
is the aggregated interference originated from \gls{PPP} distributed interfering \gls{UE}s with density $\lambda$, all signals generated by all \gls{UE}s are $x_i\sim CN(0,1)$. We assume that the aggregate interference power $I=|\iota|^2$ is typically $\mathbb{E}[I]\gg\sigma_{w}^{2}/N$ as macro-cell network \cite{HeathUplinkSINRRateSG}. This assumption is relaxed in Sect. \ref{sec:NLOS}, since in high frequencies, the noise is not negligible in small cells, specially in presence of blockage \cite{NoiseLimitedBlockage}.

The outage analysis depends on the \gls{PDF} of the aggregated interference
(\ref{eq:LOSInterfAmplitude}) for \gls{PPP} distribution of \gls{UEs} having
each polar coordinates $(R_{i},\theta_{i})$.
The \gls{PDF} of $I$  without any array-gain for interference mitigation (here is a specific case with $G^{2}(\phi_{i},\theta_{i})=1$ for any $i$) and $h=0$ has been extensively investigated in the literature (see e.g, \cite{WinPintoMatThInterference,SuuccessiveInterfCancellation,CognitiveNetworkInterf}).
The scope here is to evaluate the \gls{PDF} of aggregated interference
$I$ for \gls{UcylA} and $h\ge 0$. The $N_{c}\times N_{v}$ \gls{UcylA} is the most general case as \gls{UCA} is when $ N_{v}=1$ and the single antenna is when $N_{c}= N_{v} =1$, as
sketched in Table \ref{tab:Array-configurations-derived}, with the corresponding references or Sections for the analytic form of \gls{CF}. Notice that the density $\lambda$ refers to the number of \textit{active} UEs per square meters coexisting on the same time-frequency and, depending on the specific radio resource allocation strategies, is likely to be meaningfully lower than effective crowd density \cite{chen2020massiveaccess}.

\begin{table}[t!]
\begin{centering}
\begin{tabular}{|c|c|c|c|c|c|}
\cline{2-6} \cline{3-6} \cline{4-6} \cline{5-6} \cline{6-6}  
\noalign{\vskip0.1cm}
\multicolumn{1}{c|}{} & \gls{UcylA} &  \gls{UCA} $h>0$ & \gls{UCA} $h=0$  &
Isotropic $h>0$ & Isotropic $h=0$ 
\tabularnewline[\doublerulesep]
\cline{2-6} \cline{3-6} \cline{4-6} \cline{5-6} \cline{6-6} 
\noalign{\vskip\doublerulesep}
\noalign{\vskip0.1cm}
\multicolumn{1}{c|}{} &
\includegraphics[width=2cm,height=2cm]{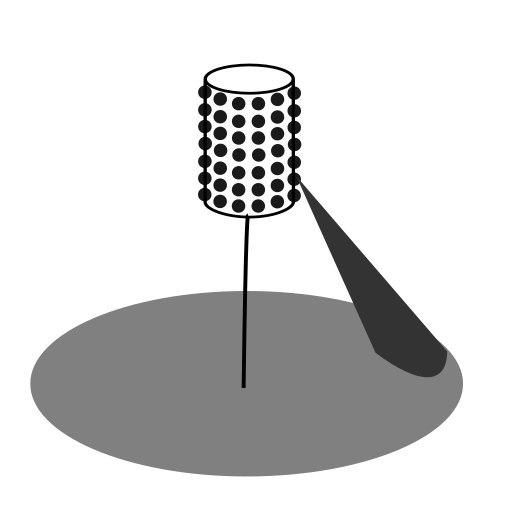} &
\includegraphics[width=2cm,height=2cm]{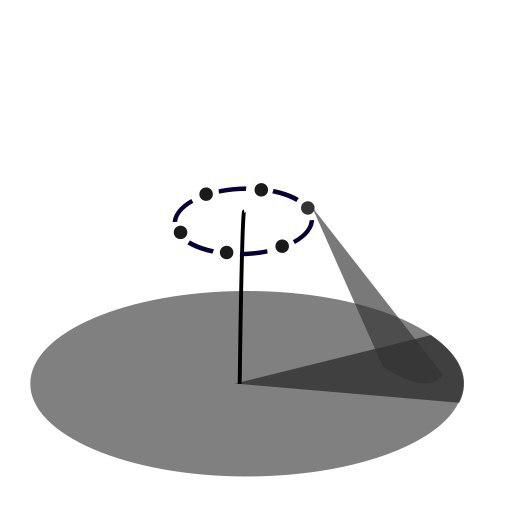} & 
\includegraphics[width=2cm,height=2cm]{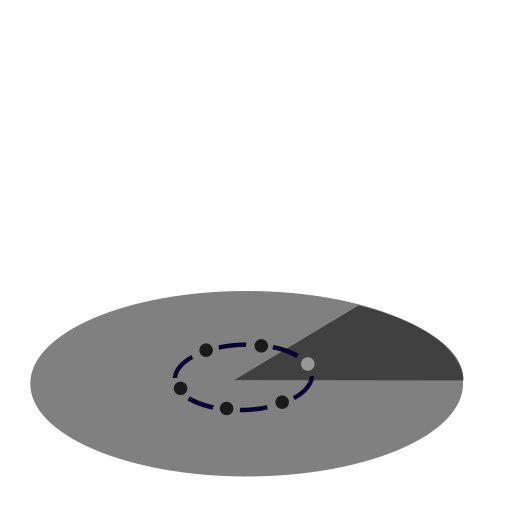} &
\includegraphics[width=2cm,height=2cm]{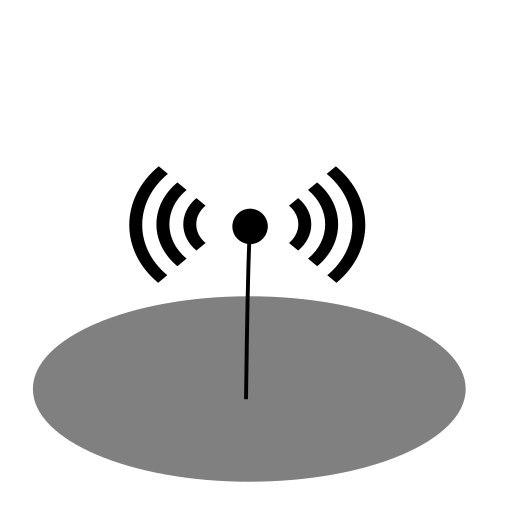} &
\includegraphics[width=2cm,height=2cm]{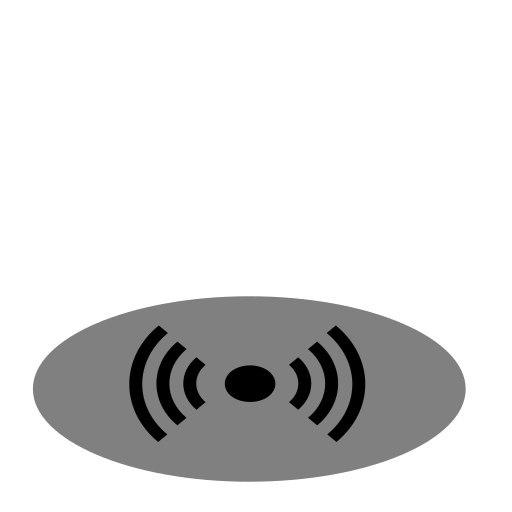}  \tabularnewline[\doublerulesep]
\noalign{\vskip\doublerulesep}
\hline 
$N_{c}$ & >1 & >1 & >1 & 1 & 1 \tabularnewline
\hline 
$N_{v}$ & >1 & 1 & 1 & 1 & 1 \tabularnewline
\hline 
$h$ & $\ge$ 0 &$\ge$0 & 0 &$\ge$0 & 0 \tabularnewline
\hline 
& %
\begin{tabular}{|c|}
\hline 
Sect.\ref{subsec:UcylA}\tabularnewline
\hline 
\hline 
\gls{CF}: (\ref{eq:CFUcylA})\tabularnewline
\hline 
\end{tabular} 
  & %
\begin{tabular}{|c|}
\hline 
Sect.\ref{subsec:SpecificCases}\tabularnewline
\hline 
\hline 
\gls{CF}: (\ref{eq:CFUCA_h>0})\tabularnewline
\hline 
\end{tabular}; & %
\begin{tabular}{|c|}
\hline 
Sect.\ref{subsec:SpecificCases}\tabularnewline
\hline 
\hline 
\gls{CF}: (\ref{eq:UCA_h=0})\tabularnewline
\hline 
\end{tabular} & %
\begin{tabular}{|c|}
\hline 
Sect.\ref{subsec:SpecificCases}\tabularnewline
\hline 
\hline 
\gls{CF}: (\ref{eq:CFpoint_antenna_h>0})\tabularnewline
\hline 
\end{tabular}
& Ref \cite{WinPintoMatThInterference}
\tabularnewline
\hline 
\end{tabular}
\par\end{centering}
\begin{centering}
\bigskip{}
\par\end{centering}
\caption{\label{tab:Array-configurations-derived}Array configurations and reference to the characteristic function (CF) of aggregated interference
$\Psi(\omega)=\mathbb{E}[e^{j\omega I}]$.}
\end{table}

The average probability of successful connection experienced by the
\gls{UE} of interest in $(\phi_{o},R_{o})$ depends on a certain threshold
$T$, on fading fluctuation $|\beta_{o}|^{2}$,
and on the overall interference $I$. The service probability from the distribution of $I$
\begin{equation}
{P}_{s}(R_{o}|R_{max})=F_{I}\left(\frac{|\beta_0|^{2}G^2(\phi_0,\theta_0)}{T{\left({R_{o}^{2}+h^{2}}\right)^{{b}}}}\right),\label{eq:Service-Probability-equation}
\end{equation}
where $F_{I}\left(x\right)=\Pr(I<x)$ is the \gls{CDF}. The service probability (\ref{eq:Service-Probability-equation}) is for the interference power $I$ which accounts for the randomness of the position of the interferers according to the PPP model within a certain radius $R_{max}$ and fading. The fluctuations of the interferers $|\beta_{i}|$ is embodied in the \gls{CF} derivations (Sect.\ref{sec:CF}). The reference user in $(R_o,\phi_o)$ is considered as deterministic for the computations of the (conditional) service probabilities, but whenever necessary for the unconditional probability, it can be assumed as PPP distributed as for the other interferers.
Notice that the service probability ${P}_{s}(R_{o}|R_{max})$ depends on the \gls{CDF} of the aggregated interference $I$ to be evaluated for the \gls{PPP} distribution of active interferers, the array-type and its height $h$, as evaluated next section from \gls{CF} analysis. For the unconditional service probability accounting for the fading of the user of interest, one should evaluate the expectation
\begin{equation}
\tilde{P}_{s}(R_{o}|R_{max})=\mathbb{E}_{|\beta_0|^{2}}\left[{P}_{s}(R_{o})\right].\label{eq:Aggregated-Service-Probability-equation}
\end{equation}
Differently from the interference analysis that aggregates multiple (and many) interfering contributions into the CF of $I$, (\ref{eq:Aggregated-Service-Probability-equation}) depends on the specific \gls{PDF} of $\beta_0$ thus making the service (or outage) analysis distribution-dependent for the UE of interest, see e.g \cite{Yongacoglu,DianaDanielaCoverageCooperation}. Therefore, the fast-fading contribution $\beta_0$ is neutrally modelled here as $\beta_0=1 \times e^{j \epsilon}$ with $\epsilon \sim \mathcal{U}(0,2\pi) $, further derivations for arbitrary fading $\beta_{0}$ are out of the scope here.

\section{\gls{CF} of the Aggregated Interference in LOS \label{sec:CF}}
In this section, it is derived the \gls{CF} of the aggregate interference in presence of LOS links first for the general case of  $N_{c}\times N_{v}$ \gls{UcylA}, then simplified for the UCA and single antenna. The focus of this paper is to infer the behaviour of the interference and coverage w.r.t. height and arrangements of the antenna array. The mmW channel modelling is first \gls{LOS}-only, then is enriched with mixed LOS and NLOS links in Sect.\ref{sec:NLOS}.
\subsection{Uniform Cylindrical Arrays\label{subsec:UcylA}} 
$N_{c}\times N_{v}$ \gls{UcylA} is composed of $N_{v}$ uniformly spaced rings consisting of $N_{c}$ antennas each arranged in a cylinder shape (Fig. \ref{fig:Cylindrical-array-configuration}). The interference power $I$ originated from a coverage radius $R_{max}\rightarrow\infty$  represents the largest possible interference for a density $\lambda$ and thus it is the upper bound of the interference $I$ when $R_{max}<\infty$, the service probability (\ref{eq:Service-Probability-equation}) depends on $R_{max}$ and it is lower bounded: ${P}_{s}(R_{o}|R_{max}) \ge {P}_{s}(R_{o}|R_{max}\rightarrow\infty)$.
 The computation of the \gls{CF} of the aggregate interference one should consider the entire gain pattern of the \gls{UcylA} (\ref{eq:directivity-patternUCA}, \ref{eq:directivity-patternULA}). Let $R_{max}\rightarrow\infty$, the aggregate interference power is
\begin{equation}
I=\sum_{i=1}^{\infty}\frac{|\beta_{i}|^{2}}{\left(R_{i}^{2}+h^{2}\right)^{b}}\thicktilde{G}^2(R_{i},\phi_{i}),\label{eq:Sum_interf_UcylA}
\end{equation}
where the served \gls{UE} is in $(\theta_{o},\phi_{o})=(0,0)$ for analytical notation convenience, and the beamforming gains are reformulated in term of azimuth ($\phi_i$) and elevation ($\theta_i=\arctan(h/R_{i})$) angles
\begin{equation}
\thicktilde{G}(R_{i},\phi_{i})=G_{c}^{2}(\phi_{i})G_{v}^{2}(\arctan(h/R_{i}))
\end{equation}
The fluctuations' power $|\beta_{i}|^{2}$ are independent of interfering users and identically distributed (iid). The randomly
distributed interfering \gls{UEs} in $\phi_{i}\in[0,2\pi)$ can be partitioned into a set of $K$ disjoint angular sectors $\Phi_{1},\Phi_{2},...$ such that $\cup_{k}\Phi_{k}\equiv[0,2\pi),$ where $K$ is large enough so that the array gain $G_{c}({\Phi}_{k})$ in each sector can be considered as constant. The aggregate interference power (\ref{eq:Sum_interf_UcylA}) is
\begin{equation}
\begin{aligned}I & =\sum_{k=1}^{K}\sum_{\phi_{i}\in\Phi_{k}}\frac{|\beta_{i}|^{2}}{\left(R_{i}^{2}+h^{2}\right)^{b}}\thicktilde{G}^{2}(R_{i},\phi_{i})\simeq\sum_{k}I_{k},\end{aligned}
\label{eq:Interf_UCA}
\end{equation}
where 
\begin{equation}
I_{k} = \sum_{i\in\Phi_{k}}\frac{|\beta_{i}|^{2}}{\left(R_{i}^{2}+h^{2}\right)^{b}}\thicktilde{G}^{2}(R_{i},{\Phi}_{k})\label{eq:Wk}.
\end{equation}
The \gls{CF} for the interference within the $k$th angular sector follows from Campbell's theorem as in \cite{WinPintoMatThInterference}
\begin{equation}
\Psi_{I_{k}}(\omega)=\exp\left(-\frac{2\pi}{K}\lambda\int_{0}^{\infty}\left[1-\Psi_{|\beta|^{2}}\left(\omega\frac{\thicktilde{G}^{2}(r,\bar{\phi}_{k})}{\left(r^{2}+h^{2}\right)^{b}}\right)\right]rdr\right)\label{eq:Campbell},
\end{equation}
where $\Psi_{|\beta|^{2}}(\omega) = \mathbb{E}[e^{j \omega |\beta|^{2} }]$ is the \gls{CF} of $|\beta|^{2}$. Defining $\omega\left(r^{2}+h^{2}\right)^{-b}=t$ and solving for
$t$ one gets (for $\alpha=1/b$)
\begin{equation}
\Psi_{I_{k}}(\omega)=\exp\left(-\frac{\pi}{K}\lambda\alpha\left|\omega\right|^{\alpha}\int_{0}^{|\omega|/h^{2b}}\dfrac{1-\mathbb{E}_{|\beta|^{2}}[e^{jt|\beta|^{2} \thicktilde{G}^{2}(f(t),{\Phi}_{k})\textrm{sign}(\omega)}]}{t^{\alpha+1}}dt\right).
\end{equation}
where $f(t)=\left(|\omega|^{1/b}t^{-1/b}-h^{2}\right)^{1/2}$ follows
from the conversion from variable $r$ to $t$. Since the array gain is a function of $t$, this expression can
only be solved numerically. One way to make the \gls{CF} tractable is
by uniformly dividing the elevation angle $\theta_{v}\in (0,\pi/2]$ into $M$ 
angular sectors of $\Delta_v=\pi/2M$ width, each sector is centered in $\thicktilde{\theta}_{m}=\frac{\pi}{2}(\frac{2(M-m)+1}{2M})$ , and the width $\Delta_v$ is small enough to let the array gain in every angular sector
(\ref{eq:directivity-patternULA}) be constant $G_{v}^{2}(\thicktilde{\theta}_{m})$. The array gain is constant on every annulus (ring) shaped areas with
unequal widths (non-uniform rings division for uniform elevations $\thicktilde{\theta}_{m}$). These rings are centered in the radial distance of the intersection of the $m$th bisector $\thicktilde{\rho}_{m}=h/\tan(\theta_{m}-\frac{\pi}{4M})$, so that the array gain in \textit{k}th wedge and \textit{m}th ring is $\thicktilde{G}^{2}(\thicktilde{\rho}_{m},{\Phi}_{k})$. Within the  \textit{k}th wedge and \textit{m}th ring the array gain is constant and the interference originated $I_{k,m}$  has the \gls{CF} 
\begin{equation}
\Psi_{I_{k,m}}(\omega) =\exp\left(-\frac{\pi}{K}\alpha\lambda|\omega|^{\alpha}\int_{\tau_{m+1}}^{\tau_{m}}\left[\frac{1-\mathbb{E}_{|\beta|^2}[e^{jt|\beta|^{2} \thicktilde{G}^{2}(\bar{\rho}_m,{\Phi}_{k})\textrm{sign}(\omega)}]}{t^{\alpha+1}}\right]dt\right),
\end{equation}
where $\tau_{m} = {|\omega|}/{(h^{2}+\rho_{m}^{2})^{b}}$. The \gls{CF} of the aggregate interference statistically independent on all $K$ wedges and $M$ rings is $\Psi_{I}(\omega)=\prod\limits_{m=0}^{M-1}\prod\limits_{k=0}^{K-1}\Psi_{I_{k,m}}(\omega)$ and it can be shown to reduce to \footnote[1]{The solution of the integral $\int_{0}^{|\omega|/h^{2b}}\left[\frac{1-e^{j \mu t}}{t^{\alpha+1}}\right]dt=\lim\limits_{\varepsilon\rightarrow0}\left.(-j\mu)^{\alpha}\Gamma(-\alpha,-j\mu t)-\frac{1}{\alpha t^{\alpha}}\right]_{t=\varepsilon}^{t=|\omega|/h^{2b}}$, for any constant and real $\mu$, and $0<\alpha < 1$.}
\begin{equation}
     \Psi_I(\omega)=\exp(-\lambda2\pi R_{max}^{2})\exp\left(-\frac{\pi\lambda}{C_{\alpha}}\left|\omega\right|^{\alpha}(1-j\textrm{sign}(\omega)\tan\frac{\pi\alpha}{2})P_{G}(\omega)\right) \label{eq:CFUcylA}
\end{equation}
with $C_{\alpha}^{-1}=\varGamma(1-\alpha)\cos(\pi\alpha/2)$ and
\begin{equation}
P_{G}(\omega)=\frac{\bar{\beta}^{2\alpha}}{K}\sum_{k=0}^{K-1}\sum_{m=0}^{M-1}\thicktilde{G}^{{2}{\alpha}}(\bar{\rho}_m,{\Phi}_{k})\left[P(-\alpha,-j|\omega|\xi(\bar{\rho}_m,{\Phi}_{k}))-P(-\alpha,-j|\omega|\xi(\bar{\rho}_{m+1},{\Phi}_{k}))\right].\label{eq:PgUcylA}
\end{equation}
We used a compact notation for different moments of fading $\bar{\beta}^{c}=\mathbb{E}[|\beta|^{c}]$, and  $P(x,z)=\int_{0}^{z}t^{x-1}e^{-t}dt/\varGamma(x)$ in (\ref{eq:PgUcylA}) is the normalized
incomplete Gamma function ratio and 
\begin{equation}
    \xi(r,\phi) = \frac{\thicktilde{G}^{2}(r,\phi)\bar{\beta}^{2}}{\left(h^{2}+r^{2}\right)^{b}}.
\end{equation}
The relationship (\ref{eq:PgUcylA}) for $M\rightarrow\infty$ and $R_{i}\leq R_{\max}<\infty$
it reduces, after some calculus, into
\begin{equation}
P_{G}(\omega)=\frac{2b}{(-j\omega)^{\alpha}\Gamma(-\alpha)}\int_{0}^{R_{max}}\left(\int_{0}^{2\pi}e^{j|\omega|{\xi}(r,\phi)}\frac{d\phi}{2\pi}\right)rdr.\label{eq:PGUcylAAssymptotic}
\end{equation}
This relation (\ref{eq:PGUcylAAssymptotic}) completes the CF of aggregated interference (\ref{eq:CFUcylA}). The \gls{PDF} and the \gls{CDF} of the interference $I$ is obtained by numeric inversion. For $h\rightarrow0$ the \gls{UcylA} degenerates into \gls{UCA}
on the ground, and the distribution of $I$ is alpha-stable for $R_{max}\rightarrow\infty$ (Sect.\ref{subsec:SpecificCases}). As seen, the general amplitude fluctuations $\beta_{i}$ of the interferers is embodied in the derivation of the \gls{CF} (whose values depend on the specific fading model chosen), however, for the numerical calculations throughout the paper, the fading power $\bar{\beta}^{2}$ is neglected because it is averaged out in the \gls{SG} over the summation of the interference for all \gls{UEs} (the fading power for the user of interest $|\beta_0|^2$ does not average out).\\
\textit{Remark 1}: Although the CF of aggregated interference (\ref{eq:CFUcylA}) for \gls{UcylA} depends on (\ref{eq:PGUcylAAssymptotic}), its numerical computation has some trade-offs. The granularity of numerical integration basically depends on the main beamwidth: the azimuth $K=N_{c}$ is a safe choice with good accuracy, while for the elevation angle $M=N_{v}/2$ to $M=N_{v}/4$ is acceptable ($M=N_{v}/2$ is a safe choice for large $\lambda$ say $\lambda > 0.1$). Thus, the complexity decrease w.r.t. massive integration, and in fact, this method would be a good way of decreasing the computation complexity. Alternatively, whenever one uses the beam gain approximation models like the flat-top model \cite{CoverageImperfectAlignmentChinese}, the summation reduces straightforwardly to two terms, and it would be quite fast in terms of computations for network analyses. 
\\\textit{Remark 2}: The array gain (\ref{eq:directivity-patternULA})
of vertical \gls{ULA} is critical for the analytical tractability of the \gls{CF} derivation, and the use of $N_{v}>1$ could be questionable for height too small. The array gain (\ref{eq:directivity-patternULA}) holds true when  the array aperture is compact compared to array height $h$ to have a plane wave-front. In practice, for \gls{mmW} communications at a frequency around (or larger) 30GHz the wavelength is approx (or smaller than ) 1cm,
and for $N_{v}=10$ antennas, the array aperture for half-wavelength antennas' spacing is $5$cm (or less). In the scenario in Fig. \ref{fig:Cylindrical-array-configuration}
the height should be above the people heights and for $h>2$m the approximation that array aperture is compact (2m$\gg$5cm) holds true.

\subsection{Specific Cases\label{subsec:SpecificCases}}
\textbf{UCA}: \gls{UCA} is a special case of \gls{UcylA} for $N_v = 1$ (i.e., $G_{v}(\theta)=1$). The statistical distribution of aggregated interference $I$ for \gls{UCA} can be adapted by considering $R_{max}\rightarrow\infty$, although it can be extended  to $R_{max}<\infty$. In \gls{UCA} there is no radial mitigation of the interference, but it is only along the azimuth. After simplifying the relation (\ref{eq:CFUcylA}) and resolving the singularity (Appendix \ref{sec:A---SingularityPoint}), it yields to:
\begin{equation}
\Psi_{I}(\omega)=\exp\left(-\frac{\pi\lambda}{C_{\alpha}}\left|\omega\right|^{\alpha}(1-j\textrm{sign}(\omega)\tan\frac{\pi\alpha}{2})P_{G_{c}(\omega)}+\pi\lambda h^{2}\right),\label{eq:CFUCA_h>0}
\end{equation}
where 
\begin{equation}
P_{G_{c}}(\omega)=\frac{\bar{\beta}^{2\alpha}}{K}\sum_{k=1}^{K}G_{c}^{2\alpha}({\Phi}_{k})P\left(-\alpha,-j|\omega|\frac{G_{c}^{2}({\Phi}_{k})\bar{\beta}^{2}}{h^{2b}}\right)\label{eq:PGUCASummation}
\end{equation}
The limit for $K\rightarrow\infty$ angular sectors is
\begin{equation}
P_{G_{c}}(\omega)=\frac{\bar{\beta}^{2\alpha}}{2\pi}\intop_{0}^{2\pi}G_{c}^{2\alpha}(\phi)P\left(-\alpha,-j|\omega|\frac{G_{c}^{2}(\phi)\bar{\beta}^{2}}{h^{2b}}\right)d\phi \label{eq:PGUCAAsymptotic}
\end{equation}
and this term has to be evaluated numerically.

Considering as special case $h\rightarrow0$, it is $P_{G_{c}}\rightarrow1$,
so the frequency dependence of $P_{G_{c}}$
vanishes, and the aggregated interference in the $k$th angular sector is skewed alpha-stable\footnote{$\mathcal{S}(\alpha,\gamma)$ denotes the skewed
stable distribution with characteristic exponent $\alpha\in(0,2]$,
unitary skewness, and scale parameter (or dispersion) $\gamma\geq0$
with a characteristic function 
\[
\mathbb{E}[e^{j\omega x}]=\begin{cases}
\exp[-\gamma|\omega|^{\alpha}(1-j\textrm{sign}(\omega)\tan\frac{\pi\alpha}{2})] & \alpha\neq1\\
\exp[-\gamma|\omega|^{\alpha}(1+j\frac{2}{\pi}\textrm{sign}(\omega)\ln|\omega|)] & \alpha=1
\end{cases}
\]}:
\begin{equation}
G_{c}^{2}({\Phi}_{k})\sum_{i\in\Phi_{k}}\frac{|\beta_{i}|^{2}}{R_{i}^{2b}}\sim\mathcal{S}\left(\alpha=\frac{1}{b},\gamma_{k}\right),
\end{equation}
where $\gamma_{k}=\frac{\Delta\phi_{k}}{2}\lambda G_{c}^{2\alpha}({\Phi}_{k})\frac{\bar{\beta}^{2\alpha}}{C_{\alpha}}$
and  $C_{\alpha}^{-1}=\varGamma(1-\alpha)\cos(\pi\alpha/2)$.
The overall interference reduces to the sum of skewed stable random terms (straightforwardly
from \cite[ eq.(1.8)]{nolan2003stable})
\begin{equation}
I=\sum_{k}I_{k}\sim\mathcal{S}\left(\alpha=\frac{1}{b},\gamma_{c}\right)\label{eq:UCA_h=0}
\end{equation}
where the total dispersion for \gls{UCA} becomes

\begin{equation}
\gamma_{c}=\pi\lambda\frac{\bar{\beta}^{2\alpha}}{C_{\alpha}}\cdot\frac{\int_{0}^{2\pi}G_{c}^{2\alpha}(\phi)d\phi}{2\pi}\label{eq:dispersionStable},
\end{equation}
assuming sectors $\Delta\phi_{k}\rightarrow0$. Therefore, the aggregated interference for UCA and $h=0$ is skewed alpha-stable. However, increasing the height $h$, the distribution deviates from alpha-stable as detailed later. 
Comparing this result with \cite{WinPintoMatThInterference} one notices
an additional term that depends on the \gls{UCA} array gain $G_{c}^{2}(\phi)$
that mitigates the mean level of interference in skewed stable distribution. As before, the fading powers of the interferers is embodied in the derivations, but they will average out in summation of the interference power over all of the UEs (the fading amplitude of each signal from each UE depends on the chosen fading model), while the fading power from the user of interest remains effective in the calculation of the service probability.

\textbf{Point antenna}: a single point antenna can be considered as a special case of a \gls{UCA}, where $N_{c}=1$ that leads to an isotropic gain. Placing the antenna at height
$h>0$ the \gls{CF} of the aggregated interference power can be achieved by simplifying (\ref{eq:CFUCA_h>0}) as
\begin{equation}
\Psi_{I}(\omega)=\exp\left(-\frac{\pi\lambda}{C_{\alpha}}\left|\omega\right|^{\alpha}\bar{\beta}^{2\alpha}P(-\alpha,-j|\omega|\frac{\bar{\beta}^{2}}{h^{2b}})(1-j\textrm{sign}(\omega)\tan\frac{\pi\alpha}{2})+\pi\lambda h^{2}\right).\label{eq:CFpoint_antenna_h>0}
\end{equation}
This \gls{CF} generalizes the \gls{CF}
for $h=0$ in \cite{WinPintoMatThInterference} and both trivially coincides for $h\rightarrow0$. The term of $P(-\alpha,-j|\omega|\frac{\bar{\beta}^{2}}{h^{2b}})$, depending on $\omega$, will increase from initial value of $P(-\alpha,-j|\omega|\frac{\bar{\beta}^{2}}{h^{2b}}) = 1$ at $h=0$.

\subsection{Analysis of \gls{AP} height}
\begin{figure}[h!]
\begin{centering}
\includegraphics[scale=0.50]{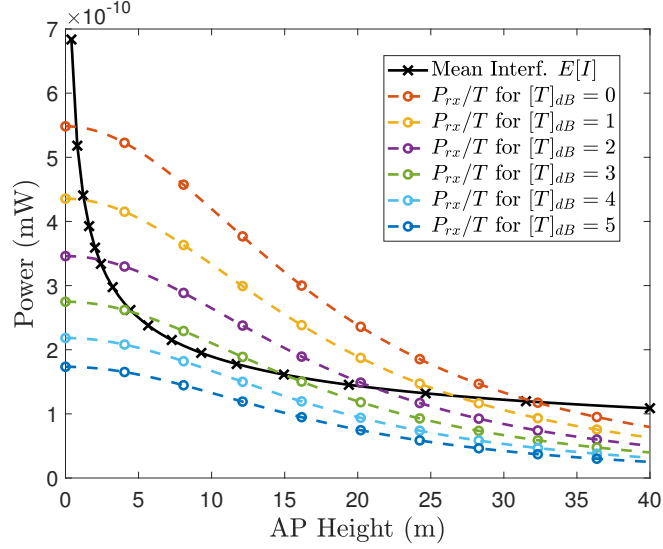}
\par\end{centering}
\caption{\label{fig:OptHeight}Received Power and mean aggregated Interference power vs. \gls{AP} height $h$, for a target UE located at $R_0 = 20$m equipped with a uniform circular array (UCA) with $N_c = 128$ that has average power gain $\bar{G_c^2}=0.012$. Parameters: SINR threshold $[T]_{dB}=\{0,1,2,3,4,5\}$, interferers' density $\lambda = 5\times10^{-3}\,m^{-2}$, path-loss exponent $2b=2.4$, central frequency $F_c = 28$ GHz, and transmit power $P_{tx} = 0 dBm$.}
\end{figure}

To gain an insight regarding the effect of the height of the antenna, it is useful to evaluate the mean aggregated interference that, for simplicity, is for UCA. The mean $ \mathbb{E}[I]$ follows from the CF properties:
\begin{equation}
    \mathbb{E}[I] = \frac{\pi\lambda \alpha^2}{(\alpha-1)}h^{2(1-b)} \, \bar{\beta^2}\bar{G_c^2}.\label{eq:MeanInterf}
\end{equation}
where $\bar{G_c^2}=\frac{1}{2\pi}\intop_{0}^{2\pi}G_{c}^{2}(\phi)d\phi$ is mean power gain. The aggregated interference power decreases as the height of the antenna increases, and increases with density $\lambda$.  As an illustrative example Fig.\ref{fig:OptHeight} shows the received signal power and the aggregated interference power case for the transmit power of all the UEs  equal $0$ dBm, and the normalized array gain for the user of interest located at distance $R_0$ is maximum (i.e. $G_c(\phi_0)=1$). The received useful signal power $P_{rx}$ by a single UE is:
\begin{equation}
    P_{rx} = \frac{|\beta_0|^2}{(R_0^2 + h^2)^b}\, ,\label{eq:UserPower}
\end{equation}
where the amplitude $\beta_o$ considers the path-loss at distance 1m: $\beta_o = 4\pi F_c/c$ for $c=10^8m/s$ and $F_c$ is the carrier frequency (here $F_c = 28$ GHz). The mean interference power follows (\ref{eq:MeanInterf}). It can be observed that for large thresholds $[T]_{dB}=\{4,5\}$ the average interference power is larger than the signal power and the system is in outage for every \gls{AP} height. For smaller thresholds, the \gls{AP} serves the target UE for a range of \gls{AP} heights. For example at $[T]_{dB}=0$, the serving range of \gls{AP} height is approximately $1m< h<33m$. At around $h=1$m, the difference of target UE signal power and mean aggregated interference is zero. For \gls{AP} height range $1m<h<9m$, this difference increases, and for \gls{AP} height range of $9m\leq h<33m$ the difference decreases, while for $33m<h$ the user is in outage.

\subsection{Numerical validation on aggregated interference\label{sec:Numerical-results}}

The \gls{CF} derived for each of the arrays is validated here by
numerically computing the \gls{PDF} and \gls{CDF} from inverse Fourier
methods tailored for statistical distributions to be accurate on the
tails of the distributions \cite{davies1973numerical,shephard1991characteristic}.
The service probability (\ref{eq:Service-Probability-equation}) is
the comparison metric adopted here for the validation of the \gls{CF}s
in the previous sections,
by considering a \gls{LOS} system with ${|\beta_o|}^{2}=1$ and threshold
$T=1$ (or $0dB$). The transmitting interferers are numerically generated as random
\gls{PPP} with a maximum radius $R_{\max}^{(\textrm{num})}$ specified below for
every Monte-Carlo iteration and are affected by the array gain $G(\phi,\theta)=G_{c}(\phi)G_{v}(\theta)$
(see (\ref{eq:directivity-patternUCA}) and (\ref{eq:directivity-patternULA}))
keeping fixed the radial position $R_{o}$ for the \gls{UE} of interest as aggregated interference is isotropic vs. azimuth. 


\begin{figure}[h!]
\begin{centering}
\includegraphics[scale=0.50]{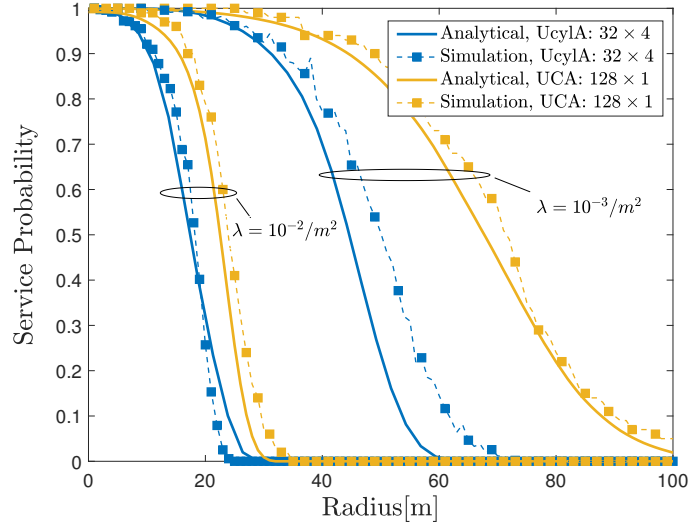} 
\par\end{centering}
\caption{\label{fig:Servprob}Service probability vs. radius $R_{o}$ for varying $\lambda=\{10^{-3},10^{-2}\}\,m^{-2}$ where the access point (AP) employs a \gls{UCA} with $N_{c}=128$ or a \gls{UcylA} with $N_c \times N_v = 32 \times 4$. Parameters: $R_{\max}^{(\textrm{num})}=200\,m$ , $[T]_{dB}=0$, $|\beta_0|^2=1$, $2b=2.6$, $h = 10\,m$.}
\end{figure}

\begin{figure}[h!]
\begin{centering}
\includegraphics[scale=0.50]{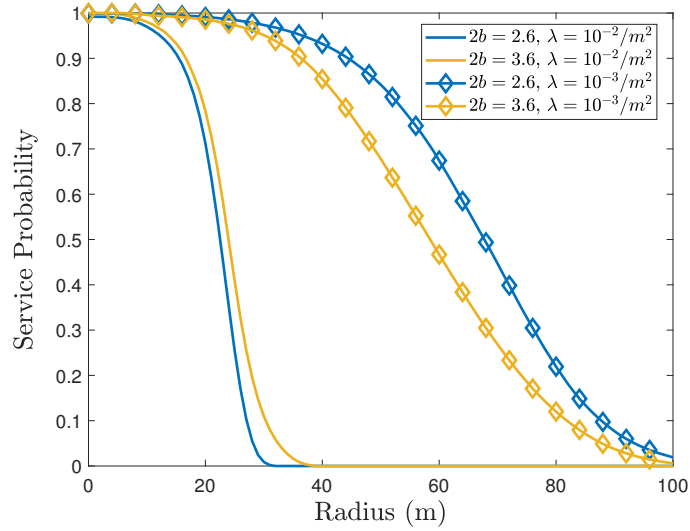} 
\par\end{centering}
\caption{\label{fig:Servprob2}Service probability vs. radius $R_{o}$ for varying $\lambda=\{10^{-2},10^{-3}\}\,m^{-2}$ where the \gls{AP} employs a \gls{UCA} with $N_{c}=128$. Parameters: $R_{\max}^{(\textrm{num})}=200\,m$ , $[T]_{dB}=0$, $|\beta_0|^2=1$, $2b=2.6$ and $2b=3.6$, $h = 10\,m$.}
\end{figure}
Fig. \ref{fig:Servprob} demonstrates the comparative analysis of \gls{UcylA} and \gls{UCA} showing the $P_{s}(R_{o})$ vs. radius $R_{o}$ using
$\Psi_{I}(\omega)$ for \gls{UcylA} in (\ref{eq:CFUcylA}) and for \gls{UCA} in (\ref{eq:CFUCA_h>0}) where the total number of antennas is preserved in all cases $(N_c \times N_v = 128)$ and the \gls{SINR} threshold $[T]_{dB} = 0$. In case of UCA, the analytical curves are a lower bound
of $P_{s}(R_{o})$ when compared to the numerical simulation with
$R_{\max}^{(\textrm{num})}\ll R_{\max}$. Simulations can show that for \gls{UCA},
increasing $R_{\max}^{(\textrm{num})}>200\,m$ (not shown here), the numerical $P_{s}(R_{o})$
attains the analytic ones.  Service probability  $P_{s}(R_{o})$ decreases for increasing served UE position $R_o$ as interference from $\lambda$-density interferers dominate. For smaller density (here $\lambda=10^{-3}\,m^{-2}$) the service probability $P_{s}(R_{o})>0.5$ up to $70\,$m For \gls{UCA} and around $50$m for UcylA. It can be noticed that in the given scenario with given parameters, the UCA seems to surpass the UcylA from service probability point of view.\\
Increasing the path-loss exponent ($2b=3.6$
in Fig. \ref{fig:Servprob2}) affects the service probability as aggregated interference is more attenuated for far away interferers, and it is more effective for denser users (i.e., for larger $\lambda$ the increase of the path-loss is more beneficial for $P_{s}(R_{o})$, while detrimental for small $\lambda$).

Once validated the analytical model, one might
investigate the $N_{c}$ vs. $N_{v}$ arrangement of \gls{UcylA} for
a given total number of antennas $N_{c}N_{v}$ (e.g., for the same complexity of the radio frequency circuitry). The cylinder arrangement of the \gls{UcylA} can be tall
($N_{c}<N_{v}$), fat ($N_{c}>N_{v}>1$) or just a ring ($N_{v}=1$)
and the optimum array geometry for service probability depends on
different parameters like the SIR threshold ($T$), path-loss exponent
($b$), the antenna (or users) height $h$, and the directivity of every
antenna element (not considered here). The metric used herein is the ratio of the served users in a certain area ($2\pi \lambda\int_{0}^{\bar{R}}P_{s}(r)rdr$) to the total average users ($\pi\bar{R}^{2} \lambda$) here assumed PPP distributed, it is also referred as average service probability
\begin{equation}
\bar{P}_{s}=\frac{2\int_{0}^{\bar{R}}P_{s}(r)rdr}{\bar{R}^{2}}.
\label{eq:averaged_Ps}
\end{equation}
In the following examples, we maintaining the total number of antennas $N_c\times N_v = 256$, changing the ratio of \gls{UCA} and vertical antennas.

\begin{figure}[h!]
\begin{centering}
\includegraphics[scale=0.35]{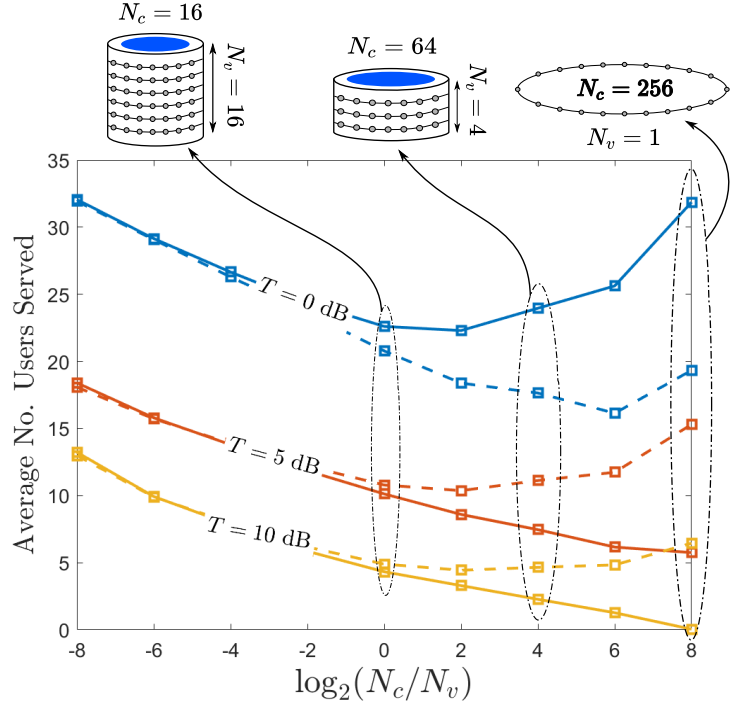}
\par\end{centering}
\caption{\label{fig:Geom_Threshold} Average service probability
$\bar{P}_{s}$ vs antennas ratio $\log_2(N_c / N_v)$, within $\bar{R}=50\,m$ while keeping constant the total number of antennas $(N_{c} N_{v}=256)$ for different thresholds $[T]_{dB} = \{0,5,10 \}$, AP height $h = 5$m, $|\beta_0|^2 = 1$, $\lambda=5 \times 10^{-2}\,m^{-2}$: Solid lines correspond to $2b = 2$ and dashed lines corresponding to $2b=3.6$.}
\end{figure}

\begin{figure}[h!]
\begin{centering}
\includegraphics[scale=0.35]{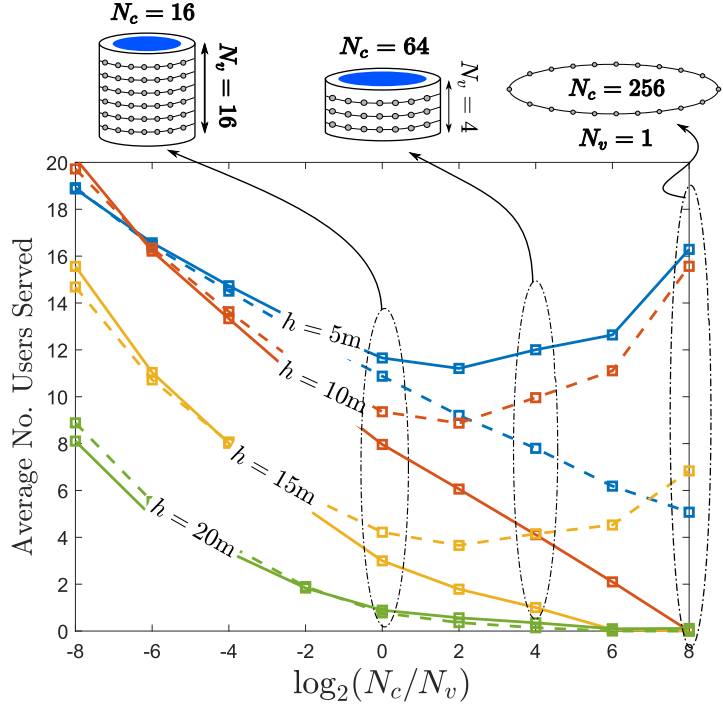}
\par\end{centering}
\caption{\label{fig:Geom_Height} Average service probability
$\bar{P}_{s}$ vs antennas ratio $\log_2(N_c / N_v)$, within $\bar{R}=100\,m$ while keeping constant the total number of antennas $(N_{c} N_{v}=256)$ for different array height $h = \{5,10,15,20\}m$, threshold $[T]_{dB}=5$,  $|\beta_0|^2 = 1$, $\lambda=5\times 10^{-2}\,m^{-2}$: Solid lines correspond to $2b = 2$ and dashed lines corresponding to $2b=3.6$.}
\end{figure}

\begin{figure}[h!]
\begin{centering}
\includegraphics[scale=0.35]{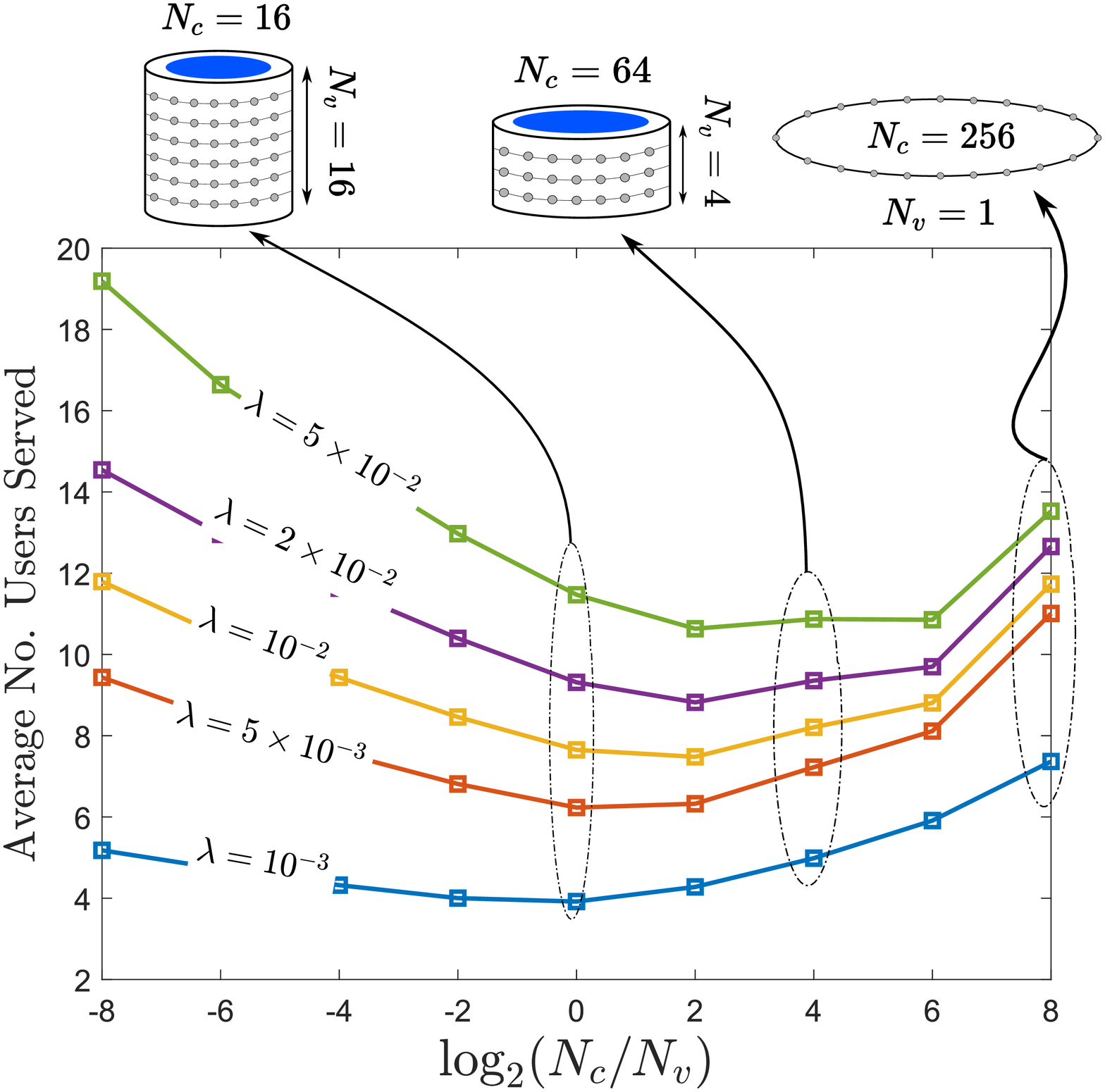}
\par\end{centering}
\caption{\label{fig:Geom_Lambda}  Average service probability
$\bar{P}_{s}$ vs antennas ratio $\log_2(N_c / N_v)$, within $\bar{R}=100\,m$ while keeping constant the total number of antennas $(N_{c} N_{v}=256)$ for different interferers' density $\lambda = \{1,5,10,20,50\}\times 10^{-3}\,m^{-2}$, threshold $[T]_{dB}=5$,  $|\beta_0|^2 = 1$ and $2b = 2$}
\end{figure}
Figure \ref{fig:Geom_Threshold} illustrates $\bar{P}_{s}$ vs. the antennas ratio $\log_2(N_c / N_v)$, varying the SINR threshold $[T]_{dB} = \{0,5,10 \}$ for small and large path-loss exponents $2b = \{2,3.6\}$. It can be seen that in the given scenario, by increasing the threshold $T$, \gls{UCA} performs worse. Note that in this setup, the UEs are very dense ($\lambda = 5 \times 10^{-2}\, m^{-2}$). It can be shown that for smaller UE densities, \gls{UCA} outperforms the \gls{UcylA}. Fig. \ref{fig:Geom_Height} shows a similar example of $\bar{P}_{s}$ vs. the antennas ratio $\log_2(N_c / N_v)$, fixing the threshold $[T]_{dB} = 5$ and varying the AP height $h = \{5,10,15,20 \}\,m$. Clearly by increasing the AP height, the usage of \gls{UcylA} becomes
more advantageous in terms of $\bar{P}_{s}$ and it is more meaningful to use vertical beamforming, while for smaller heights it is preferable to use a larger circular array. However, it can be observed that in this scenario where UEs are dense, the service probability in general decreases by increasing the height. Thus, it is preferable to use an AP with smaller height.\\ Fig. \ref{fig:Geom_Lambda} shows the effect of UEs density $\lambda$. It is seen that for areas that are not so dense, a \gls{UCA} is mildly preferable. However for more dense areas, a \gls{UcylA} would be more preferable depending on different parameters. It can be shown that changing every one of the above-mentioned parameters affects the shape of the $\bar{P}_{s}$ curves. Furthermore, the constraint of ground UEs is also responsible for results. \gls{UcylA}s are favored in case UEs have arbitrary heights (usually indoor UEs are at arbitrary heights, while in high frquencies we target, penetration loss is too high).

A pragmatic conclusion from these evaluations is that for most real-life scenarios where UEs densities are not too high, it is beneficial to invest in \gls{UCA} arrangement rather than \gls{UcylA}. There are two reasons supporting this result: i) the most powerful interferers are in the vicinity of the \gls{AP}, while the minimum separable angle by the ULA is $\Delta\theta_{min} \approx \frac{\lambda}{N_v d_{v} cos(\theta_0)}$, where $d_v$ is the vertical inter-element spacing, and $\theta_0$ is the target elevation; ii) even if the resolution was not angle dependent, dividing the whole elevation plane into small portions of same width is emphasizing the same way to near and to far UEs, while the most powerful UEs are closer ones. In the rest of the numerical examples in the paper, we focus on the usage of \gls{UCA}, since the target threshold used is set to $[T]_{dB} = 0$ and the interferers' density used are not extremely high, that justifies the usage of a \gls{UCA} over \gls{UcylA}.\\
\textit{Remark 3:} Note that although the left most parts of the curves (i.e. corresponding to a \gls{ULA} or a very tall \gls{UcylA}) are shown in the figures, in practice they are not feasible to deploy. Therefore, one might consider the range $N_c> N_v/8$ as practical solution.
\section{Statistical Characterization of the aggregated interference power\label{sec:InterfCharact}}
\begin{figure}[h!]
\begin{centering}
\includegraphics[scale=0.50]{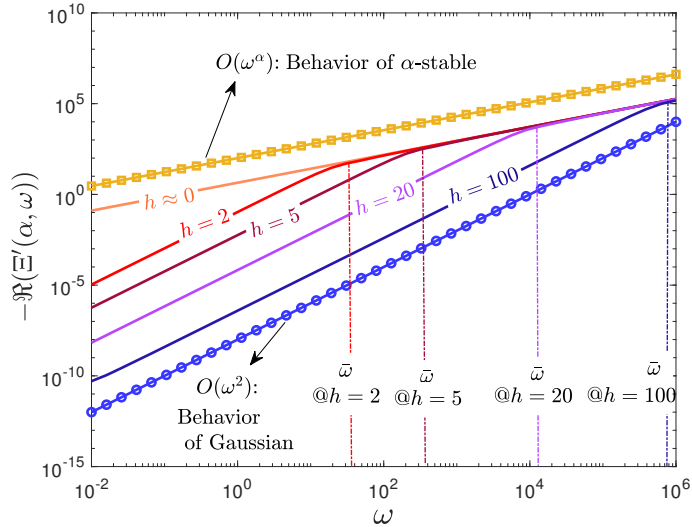}
\par\end{centering}
\caption{Real part of the ${\varXi(\alpha,\omega)}^{\prime}$ vs. $\omega$ for $2b = 2.6$ and $\lambda = m^{-2}$, with a single isotropic antenna for different heights of the array. The two guidelines are parallel with $\omega^{\alpha}$ and $\omega^{2}$. For large heights, ${\varXi^{\prime}(\alpha,\omega)}$ vs $\omega$ behaves as $\omega^{2}$, for very small heights it behaves as $\omega^{\alpha}$, while for medium heights it has two different slopes based on $\omega$.  \label{fig:OmegaDependance}} 
\end{figure}

\begin{figure}[h!]
\begin{centering}
\includegraphics[scale=0.50]{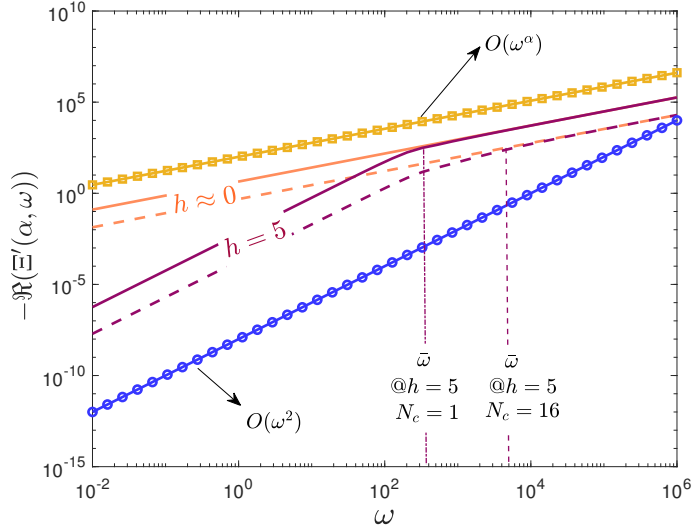}
\par\end{centering}
\caption{Real part of the ${\varXi(\alpha,\omega)}^{\prime}$ vs. $\omega$ for $2b = 2.6$  and $\lambda = 1 \, m^{-2}$, comparing the UCA with single antenna. The two guidelines are parallel with $\omega^{\alpha}$ and $\omega^{2}$: Solid lines correspond to a point antenna while dashed lines correspond to a \gls{UCA} with $N_c = 16$ isotropic antennas on a ring.  \label{fig:OmegaDependance2}}
\end{figure}

The aggregate interference $I$ in Sect.\ref{subsec:UcylA} for \gls{UcylA} and arbitrary height $h>0$ is complex to be computed or analytically in a closed-form; herein we propose a methods for statistical approximation of the aggregate interference \gls{CF}. We show that the aggregated interference power for an array of antenna located at arbitrary height, can be approximated by a weighted mixture of two stable distributions and we detail herein the equivalent CF.\\
In order to get a deeper insight into the distribution in an arbitrary height, one can start from the Taylor series of the argument of the CF ($\Psi_{I}(\omega)$). For example, the series for a \gls{UCA} with arbitrary height and $|\beta_{i}|=1$, follows from the CF (\ref{eq:CFUCA_h>0}) that, with some simplifications, yields:
 \begin{equation} 
\Psi_{I}(\omega) =\exp\left({\pi \alpha \lambda}{\varXi(\alpha,\omega)}\right),\label{eq:Expansion2}
\end{equation}
where the Taylor series is
\begin{equation}
    \varXi(\alpha,\omega) =\frac{1}{K} \sum_{k=1}^{K}\sum \limits_{i=1}^{\infty} \frac{(\frac{1}{h^{2b}})^{i-\alpha}}{(i-\alpha)i!}\left(j\omega\bar{\beta}^{2}\,G_{c}^{2}({\Phi}_{k})\right)^{i}\label{eq:CF_Series}.
\end{equation}
Let's define $G_{avg}^{(z)} = \frac{1}{K}\sum_{k=1}^{K}G_{c}^{2\,z}({\Phi}_{k})$, for $h=0$ relation (\ref{eq:CF_Series}) simplifies to
\begin{equation}
 \varXi(\alpha,\omega)=(-j\omega)^{\alpha}\Gamma(-\alpha)\bar{\beta}^{2\alpha}\,G_{avg}^{(\alpha)}.   
\end{equation}
For very large heights $h\gg0$ in (\ref{eq:Expansion2}), the terms with higher index $i$ get negligible, and it can approximated with only the first two terms, which would make the \gls{CF} ($\Psi_{I}(\omega)$) be a Gaussian distribution\footnote[3]{The CF of a Gaussian distribution $\exp\left(j\mu\omega-\frac{\sigma^{2}}{2}\omega^{2}\right)$ with shift $\mu$.}. In order to evaluate the appropriate \gls{CF} for any $h>0$ one isolates the behaviour vs. $j\omega$ from $\varXi(\alpha,\omega)$ as:
\begin{equation}
    \varXi(\alpha,\omega) = {\varXi^{\prime}(\alpha,\omega)}  + \frac{(\frac{1}{h^{2b}})^{1-\alpha}}{1-\alpha}j\omega\bar{\beta}^{2}\,G_{avg}^{(1)}\label{eq:VarXiPrimeFormula},
\end{equation}
where the second term (corresponding to $i=1$ in (\ref{eq:Expansion2})) is a shift or location parameters. The behavior of the real part of $\varXi(\alpha,\omega)^{\prime}$ reveals the corresponding exponent of $\omega$, for every defined $\omega$. The reason to separate ${\varXi(\alpha,\omega)}$ in two parts is that we need to omit the shift, to be able to visualise the exponent of the $\omega$ within the distribution. For large h,  ${\varXi^{\prime}(\alpha,\omega)}$ vs $\omega$ behaves as $\omega^{2}$, for very small h behaves as $\omega^{\alpha}$, while for medium heights it has two different slopes based on $\omega$. The transition $\omega$ where the behaviour changes is $\bar{\omega}$, that depends on the height, the path-loss exponent, UEs density and array gain. Having gained insight regarding the behavior vs $\omega$ and $\bar{\omega}$ which is explained in further text, ${\varXi(\alpha,\omega)}$ can be approximated as:
\begin{equation}
    \varXi(\alpha,\omega) = W_{1}(\omega)\times  a(\omega) + W_{2}(\omega)\times \left( 1-a(\omega)
\right),\label{eq:MixtureDist} 
\end{equation}
where $a(\omega)$ is a Heaviside step function, i.e., $a(\omega)=1$ for $\omega<\bar{\omega}$ and $0$ otherwise (or some function with smoother transition), that acts as a switch between two cases with different behaviour:
\begin{eqnarray}
W_{1}(\omega) & \approx &   \frac{(\frac{1}{h^{2b}})^{1-\alpha}}{1-\alpha}j\omega\bar{\beta}^{2}\,G_{avg}^{(1)}-\frac{(\frac{1}{h^{2b}})^{2-\alpha}}{2(2-\alpha)}\omega^{2}\bar{\beta}^{4}\,G_{avg}^{(2)}\label{eq:WeightGaussian}\\
W_{2}(\omega) & = & (-j\omega)^{\alpha}\Gamma(-\alpha)\bar{\beta}^{2\alpha}\,G_{avg}^{(\alpha)},\label{eq:Weight}
\end{eqnarray}
where $W_2$ coincides with the skewed-stable distribution (\ref{eq:UCA_h=0}). Relation (\ref{eq:MixtureDist}) means that the CF is decomposable as
 \begin{equation} 
\Psi_{I}(\omega) = \exp\left(\pi \alpha \lambda \,W_{1}(\omega)\,  a(\omega)\right)\,.\,\exp\left(\pi \alpha \lambda \,W_{2}(\omega)\, (1-a(\omega))\right) .\label{eq:DecomposableCF}
\end{equation}
Fig. \ref{fig:OmegaDependance} demonstrates the behavior of the real part of $\varXi(\alpha,\omega)$ vs $\omega$ for a single isotropic antenna and set of AP heights $h = \{0,2,5,20,100 \}m$. It can be seen that by increasing the \gls{AP} height, the breaking frequency $\bar{\omega}$ increases, while for extremely large \gls{AP} heights, it tends to infinite that is Gaussian distribution behavior. Fig. \ref{fig:OmegaDependance2} is the same analysis, comparing a isotropic antenna with a \gls{UCA} consisting of $N_c = 16$ isotropic antennas for the set of \gls{AP} heights $h = \{0,5 \}m$. It is observed that the slopes of the curves are maintained, while the breaking frequency is increased when using a \gls{UCA}.

One can approximate the CF for a UCA in every arbitrary height, given the knowledge about the transition frequency $\bar{\omega}$.
Knowing the transition point $\bar{\omega}$, based on height $h$ one may characterize with different statistical distributions for different $\omega$. Numerical formulation of $\bar{\omega}$ vs. antenna height and other parameters can be investigated further. A rule of thumb for  $\bar{\omega}$ can be achieved as follows. Let $S_i(\omega)$ denote the $i$-th term of the series (\ref{eq:CF_Series}) as:
\begin{equation}
    S_i(\omega) = \frac{(\frac{1}{h^{2b}})^{i-\alpha}}{(i-\alpha)i!}\left(j\omega\bar{\beta}^{2}\,G_{avg}^{(i)}\right)^{i}.
\end{equation}

It is empirically observed that $\bar{\omega}$ can be approximately achieved by imposing the condition $S_2({\omega}) = \, S_3({\omega})/\bar{\beta}^{2}$ and by solving for $\omega$ and the $\bar{\omega}$ can be achieved. Please note that this formula holds for normalized array gain. The rationale behind these conditions lies in the fact that for an $\alpha$-stable distribution and for height $h=0$ all the terms for $i>1$ tend to infinite, while for Gaussian distribution only the first three terms exist. In the mixture case, for $\bar{\omega}<\omega$ behaviour resembles $\alpha$-stable. In the case of Fig. \ref{fig:OmegaDependance} the calculated $\bar{\omega}$ for the \gls{AP} heights of $h = \{2,5,10,100\}$m are respectively $\bar{\omega} = \{32, 355, 13070, 858160 \}$, and in Fig. \ref{fig:OmegaDependance2}, the $\bar{\omega}$ for \gls{AP} height $h=5$m for two cases of $N_c = \{1,16\}$ are respectively $\bar{\omega} = \{355, 5688\}.$ Based on the figures, it can be noticed that these approximation are close to the real breaking points of the curves.

In practical systems, the height is known but other parameters such as the density of active users $\lambda$ is not known, and some inaccuracies w.r.t. the ideal model might occur. We believe that the knowledge of a reasonable approximation of the distribution of the aggregated interference enables the possibility to measure the approximating alpha-stable distribution during multiple idle times of the communication intervals by any unsupervised learning method \cite{UnsupervisedLearning} being a practical on-the-fly method.

\section{Outage Analysis in presence of NLOS, Noise and Blockage  \label{sec:NLOS}}
In the previous sections, we derived the CF of the aggregated interference power with different antenna array configurations, and characterized the distribution of the aggregated interference power. Moving toward the modelling of practical \gls{mmW} and sub-THz systems, in this section, it is shown how the NLOS propagation, noise power and blockage can be integrated into the model.
\subsection{NLOS paths}
Previous studies and measurement campaigns \cite{NumberOfNLOsPaths}, have shown that the NLOS clusters of rays are present in \gls{mmW} communications where they form sparse multipath faded channels. These paths can increase the amount of interference, but at the same time would lead to more useful signal received in the case of coherent reception of signal. On the other hand, diversity is an efficient way to compensate the blockage and to increase the reliability of communication systems.
\subsection{NLOS paths}
\begin{figure}[t!]
\begin{centering}
\includegraphics[scale=0.5]{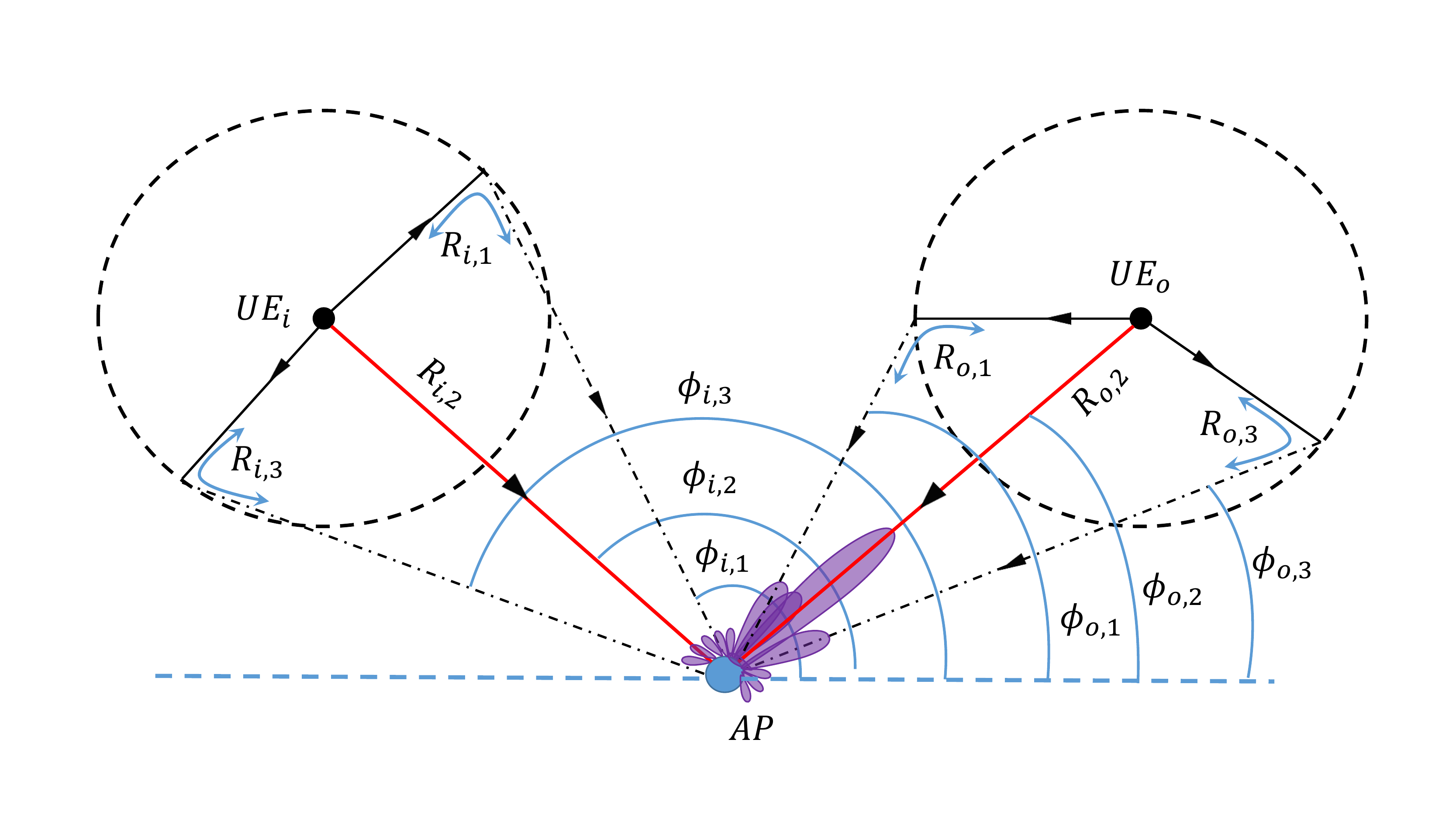}
\par\end{centering}
\caption{NLOS model: every user has few NLOS links in addition to a LOS link. LOS links are shown with thick red lines, and NLOS links are shown with dashed lines reflected back from the perimeter of a circle around the UE. Here the $UE_o$ is the user of interest and $UE_i$ is an interferer.\label{fig:NLOSModel}}
\end{figure}
Let $L$ be the total number of paths that the signal arrives from the user of interest to the same \gls{AP}, the $L\times1$
set of signals $\mathbf{y}=[y_{1},y_{2},...,y_{L}]^{T}$ after the multi-beam beamforming to each of the paths from the user of interest transmitting $x$ is
\begin{equation}
\mathbf{y}=\mathbf{h}x+{\boldsymbol{\iota}}+{\mathbf{w}}\label{eq:ReceivedSignalMR}
\end{equation}
where $[\mathbf{h}]_{\ell}=h_{\ell}=\beta_{o,\ell}/D_{\ell}^{b}$ for distance $D_{\ell}=(R_{o,\ell}^{2}+h^{2})^{1/2}$ corresponding to the $\ell$th paths of arrival (in case  $\ell=1$ it is the direct LOS link and $R_{o,1}=R_o$ is the geometric distance between the user of interest, while the distances for NLOS links are modelled later but  $R_{o,\ell}\ge R_o$. The ensemble of the aggregated interference amplitudes from the PPP distributed interferers is ${\boldsymbol{\iota}} = [{\iota}_{1},{\iota}_{2},...,{\iota}_{L}]^{T}$ that are independent and identically distributed (iid) random variables are obtained from a set of $L$ beamforming toward the distinct angles $\phi_{o,1},\phi_{o,2},...,\phi_{o,L}$ for LOS ($\phi_{o,1}$) and NLOS ($\phi_{o,2},...,\phi_{o,L}$) of the user of interest, so that adapting \ref{eq:LOSInterfAmplitude} to this case with multipath for interference is:
\begin{equation}
    \iota = \sum_{i=1}^{\infty}\sum_{\ell=1}^{L}\frac{\beta_{i,\ell}}{\left({R_{i,\ell}^{2}+h^{2}}\right)^{\frac{b}{2}}}G_{c}(\phi_{i,\ell})x_{i,\ell}\label{eq:NLOSInterfAmplitude}.
\end{equation}
This assumption is justified by the interfering ray-paths on every beamforming that have different attenuations and phase shifts, thus independent. $\mathbf{w}=[{w}_{1},{w}_{2},...,{w}_{L}]^{T}$ is the collection of noise amplitudes. The LOS/NLOS links are shown in Fig. \ref{fig:NLOSModel} and using the Weyl model (similar to Saleh-Valenzuela \cite{SalehValenzuela} adapted for mmW \cite{ValenzuelammWAdaptation,ValenzuelammWAdaptation2}) where the indirect NLOS paths from the transmitter are reflected from a secondary point which is uniformly distributed around a circle with radius $d$ around each of the transmitter's location. Usually, at high frequencies, there are not many NLOS paths so that typically are $L=2-3$ \cite{RapaportNLOSmodel,RapaportWidebandNLOSmodel}, \cite{mmWChannelModel_Rappaport}. 
The receiver for $L$ paths, possibly with different (and likely delay-resolvable for mmW and sub-THz system with large bandwidth) delays is expected to combine to maximize the service probability.
If using the \gls{MRC} combiner for the $L$ paths related to the user of interest affected by the interference powers $I_\ell =|{\iota}_{\ell}|^2$, one gets the following service probability analysis (see Appendix \ref{sec:AppendixMRC} for derivation and specific \gls{MRC} notation):
\begin{equation}
    {P}_{MRC} = \mathbb{P}_{I} \left(\sum_{\ell=1}^{L}{\alpha_{\ell}{I}_{\ell}} < \frac{1}{T}   \left( \sum_{\ell=1}^{L}\frac{|\beta_{\ell}|^2}{\bar{{I}}_{\ell}^{2 }\, D_{\ell}^{2b}}  \right)^{2}  \right)\label{eq:PserviceNLOSMRC}.
\end{equation}
where $|\beta_{\ell}|^2$ is the fluctuation for the signal from the user of interest from $\ell$th path, and the distribution of total interference $I_{MRC}=\sum_{\ell=1}^{L}{\alpha_{\ell}{I}_{\ell}}$ follows from the \gls{CF} for iid interferers over the L-beamformers:
\begin{equation}
    \Psi_{{I_{MRC}}}(\omega) = \prod_{\ell=1}^{L}\Psi_{{I}}(a_{\ell}\omega)\label{eq:CFNLOSMRC}.
\end{equation}

The NLOS distances $D_{\ell}=(R_{o,\ell}^{2}+h^{2})^{1/2}$, and $R_{o,\ell}$ for $\ell>1 $ are affected by the random angular position $\eta$ of the NLOS reflections around the radius $d$ as depicted in Fig. \ref{fig:NLOSModel}: 
\begin{equation}
R_{o,\ell}  =   d+\sqrt{R_{o,1}^2 + d^2 - 2 \, d \, R_{o,1}\cos(\eta)} \; \forall \; \ell >1 .\label{eq:DistanceNLOS}
\end{equation}
Pragmatically, NLOS distance for the served user is dependent on the specific multipath model assumed here, and a convenient way to incorporate the NLOS attenuation from the additional path $R_{o,\ell}  -R_{o,1}$ is to approximate this term is by considering the mean distance for NLOS  $\bar R_{o,\ell}=E_\eta [R_{o,\ell}]$. The distance $\bar D_{\ell}=(\bar R_{o,\ell}^{2}+h^{2})^{1/2}$ is for the NLOS  $\ell>1 $ and thus the relationship (\ref{eq:PserviceNLOSMRC}) is somewhat simplified by constant distances. Recalling that LOS/NLOS models hold for interferers' $i_{\ell}$; the power is augmented by the NLOS components, and service probability is reduced accordingly compared to $L=1$ (LOS-only). The CF for $L>1$ is derived in Appendix \ref{sec:AugmentedInterference}, accounting for LOS and NLOS.
\begin{figure}[h!]
\begin{centering}
\includegraphics[scale=0.50]{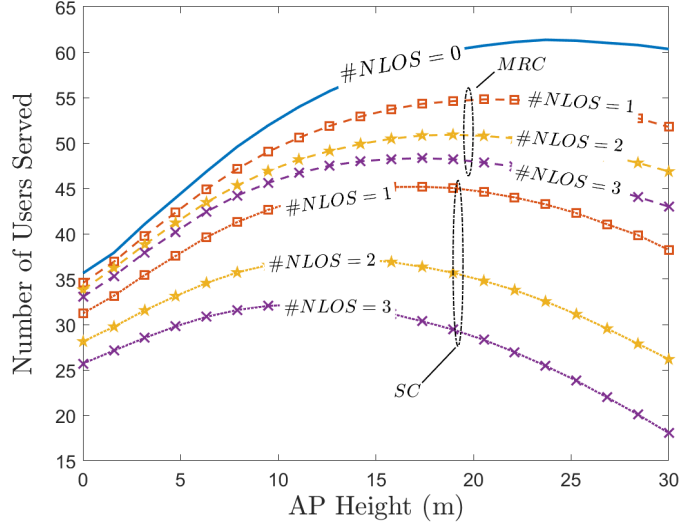}
\par\end{centering}
\caption{\label{fig:NLOSMRCSC1}Average users served $M_{s}$ within square area $200\,m \times 200\,m$ vs. \gls{UCA} height $h$, solid lines is the no NLOS ($L=1$), while dashed lines with marker are with NLOS for $L=2,3,4$, with maximal ratio combining (MRC) and selection combining (SC) receivers for $N_c = 500, \lambda=10^{-2}\, m^{-2}$, $2b = 2.6$, $[T]_{dB} = 0$, $|\beta_{\ell}|^2=1$ for every path $\ell$}
\end{figure}

\begin{figure}[h!]
\begin{centering}
\includegraphics[scale=0.50]{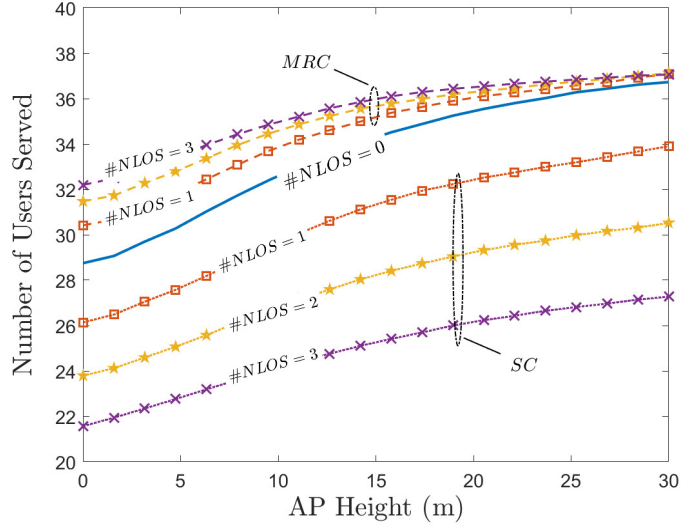}
\par\end{centering}
\caption{\label{fig:NLOSMRCSC2} Average users served $M_{s}$ within square area $200\,m \times 200\,m$ vs. \gls{UCA} height $h$, solid lines is the no NLOS ($L=1$), while dashed lines with marker are with NLOS for $L=2,3,4$, with MRC and SC receivers for $N_c = 500, \lambda=10^{-3}\, m^{-2}$; $2b = 2.6$, $[T]_{dB} = 0$, $|\beta_{\ell}|^2=1$ for every path $\ell$ }
\end{figure}

Fig. \ref{fig:NLOSMRCSC1} and Fig. \ref{fig:NLOSMRCSC2}  show the average number of users served $M_s$ within a $\Delta \times \Delta$ square-shaped area (from (\ref{eq:averaged_Ps}) it is $M_s=\lambda \Delta^2 \bar{P}_s$) when the \gls{AP} with \gls{UCA} is located at the center. The multipath increase the interference and increasing the number of NLOS paths from 1 to 3 (or $L=1,2,3,4$).
In Fig. \ref{fig:NLOSMRCSC1} the performance degrades as the density of the interferers are quite high (here $\lambda=10^{-2}\,m^{-2}$), but differently in Fig. \ref{fig:NLOSMRCSC2} the aggregated interference is lower due to the lower density ($\lambda=10^{-3}\,m^{-2}$) and thus there is a clear benefit arising from the multipath that vanishes for large height (here $h>25m$). The performance from MRC received in Fig. \ref{fig:NLOSMRCSC1} and Fig. \ref{fig:NLOSMRCSC2} is compared to the selection combining (SC) that select the path with the largest SIR \cite[Ch.\ 7]{goldsmith}. Derivation of SC is straightforward (not shown here). As expected, the MRC outperforms the SC, but one might notice that degradation becomes more severe for large multipath. In the remainder of the paper, we consider the multipath condition with both \gls{LOS} and \gls{NLOS} links.

\subsection{Noise\label{subsec:Noise}}
Even if the paper analyzes the aggregated interference power, in a 6G network with high path-loss it is inevitable to consider also the effect of the noise power. To take the noise power account, it is easy to prove that relation (\ref{eq:PserviceNLOSMRC}) must be modified:
\begin{equation}
    \mathbb{P}_{service} = \mathbb{P}_{I} \left(I_{tot}+  N_{tot} < \frac{1}{T}   \left( \sum_{\ell=1}^{L}\frac{|\beta_{\ell}|^2}{\bar{I}_{\ell}\, D_{\ell}^{2b}}  \right)^{2} \right)\label{eq:PservieConsiderNoise},
\end{equation}
where $I_{tot}=\sum\limits_{\ell=1}^{L}{{\alpha}_{\ell} {I}_{\ell}}$ and  $N_{tot}=\sigma_{n}^{2}\sum\limits_{\ell=1}^{L}{\alpha_{\ell}}$ are the total interference power and noise power respectively, after \gls{MRC} at the \gls{AP} with $\alpha_{\ell}=|\beta_{\ell}|^2/{(\bar{I}_{\ell}+\sigma_{n}^{2})^{2 } D_{\ell}^{2b}}$.

\subsection{Blockage\label{subsec:Blockage}}
\begin{figure}[h!]
\begin{centering}
\includegraphics[scale=0.50]{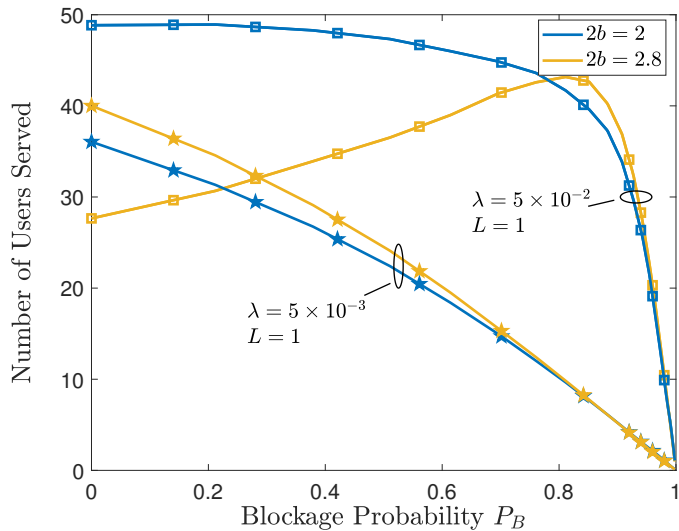}
\par\end{centering}
\caption{\label{fig:Us_Vs_Pb} Average service probability vs the probability or blockage of all the links, for $\lambda=\{5\times10^{-3},5\times10^{-2}\}\, m^{-2}$, and number of paths $L=\{ 1,4\}$ where $L=1$ means that only LOS link exists.  Parameters: $P_{tx}=20$ dB, $NF = 7$dB, $BW=400$ MHz, $F_{c} = 28$ GHz, service area = $100m \times 100m$ square, \gls{AP} height $h=10$m, threshold $[T]_{dB} = 0$. }
\end{figure}

In mmW and 6G sub-THz systems the waves are prone to blockage due to static and dynamic blockage. Static blockage \cite{StaticBlockage1,StaticBlockage2,Blockage1} is caused by  structures like buildings, trees, etc., self-blockage \cite{MacroDiv_SelfBlockage,Blockage1,Indoor_CircularBlockage} is caused by the body holding the UE, and dynamic blockage \cite{Blockage1} is caused by moving objects, humans or vehicles. The behavior of blockages and their impact on the coverage probability and system performance are different. Some of the blockage effects can be modeled in closed form, while some others need numerical methods. Here we limit the analysis of the paper to the numerical evaluation of the coverage probability in the presence of blockage. \\
It is convenient to define a dummy binary variable $\mu_{\ell}$ for the $\ell$-th path, which is $\mu_{\ell}=0$ when the link is blocked and $\mu_{\ell}=1$ when the link is free of any blockage, such that $Prob(\mu_{\ell}=0)=P_B$ is invariant on every path, and $P_B$ is the probability of blockage. The blockage per path can be incorporated into (\ref{eq:PservieConsiderNoise}) as
 \begin{equation}
    \mathbb{P}_{service} = \mathbb{P}_{I} \left(\tilde{I}_{tot}+ \tilde{N}_{tot} < \frac{1}{T}   \left( \sum_{\ell=1}^{L}\frac{|\beta_{\ell}|^2\mu_{\ell}}{\bar{I}_{\ell}\, D_{\ell}^{2b}}  \right)^{2} \right)\label{eq:ServiceDynamicBlockage},
\end{equation}
where $\tilde{I}_{tot}=\sum\limits_{\ell=1}^{L}{{\tilde{\alpha}}_{\ell} {I}_{\ell}}\,\mu_{\ell}$ and $\tilde{N}_{tot}=\sigma_{n}^{2}\sum\limits_{\ell=1}^{L}{\tilde{\alpha}_{\ell}}$ are the total interference and noise power respectively after \gls{MRC} at the \gls{AP}, with $\tilde{\alpha}_{\ell}=\mu_{\ell}\alpha_{\ell}$ being a scaling parameter for a given path, and it is obviously switched off by $\mu_{\ell}$ when the path is blocked.\\  
The goal is to assess the effect of the impact of blockage $P_B$ on  service probability. At each snapshot, one link can be either blocked or available. In order to calculate the service probability, taking into account the blockage probability, one must numerically evaluate (\ref{eq:ServiceDynamicBlockage}), for $Prob(\mu_{\ell}=0)=P_B$ for $\forall \ell$. 
\begin{figure}[h!]
\begin{centering}
\includegraphics[scale=0.50]{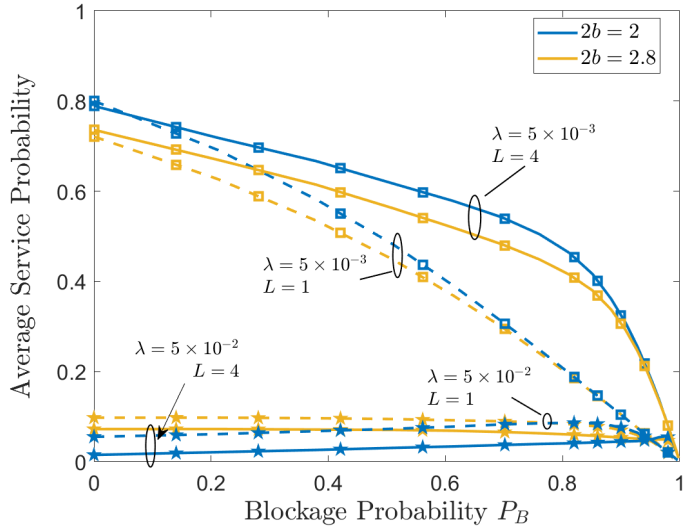}
\par\end{centering}
\caption{\label{fig:Ps_Vs_Pb}Average service probability vs the probability or blockage of all the links, for $\lambda=\{5\times10^{-3},5\times10^{-2}\}\, m^{-2}$, and number of paths $L=\{ 1,4\}$ where $L=1$ means that only LOS link exists. Parameters: $P_{tx}=20$ dB, $NF = 7$dB, $BW=400$ MHz, $F_{c} = 28$ GHz, service area = $100m \times 100m$ square, \gls{AP} height $h=10$m, threshold $[T]_{dB} = 0$.}
\end{figure}
Fig. \ref{fig:Ps_Vs_Pb} is the average service probability and Fig. \ref{fig:Us_Vs_Pb} is the number of UEs served, both figures are versus blockage probability  $P_B$ for small ($\lambda=5\times10^{-3}\,m^{-2}$)  and large ($\lambda=5\times10^{-2}\,m^{-2}$) UEs density, and varying number of paths ($L={1,4}$), and path loss ($2b=2,2.8$). 
On Fig. \ref{fig:Ps_Vs_Pb}, for small $\lambda$ and LOS path ($L=1$) the blockage probability makes the the service probability drop, while for $L=4$ the service probability is more robust even for large $P_B$, and this is due to the diversity of the multipath against the interference.
On the other hand, when $\lambda =5\times10^{-2}\,m^{-2}$ a multipath channel with $L=4$ degrades severely the performance. Numerical analysis shows the impact of the blockage on the service probability  in Fig.\ref{fig:Ps_Vs_Pb}, and the average number of user served in Fig. \ref{fig:Us_Vs_Pb}. The total number of users within a $\Delta \times \Delta$ square-shaped area is $M_s=\lambda \Delta^2\bar{P}_s$. This means that for small $P_B$ any increase of one decade of UE density makes a smaller variation in the number of users served. Recall that the number of UEs served in Fig. \ref{fig:Us_Vs_Pb} refers to the UEs allocated in the same spectrum, and thus knowing the average number of UEs assignable on the same spectrum region, one can pre-design the largest number of users that a resource scheduler can expect to assign (not covered in this paper). 
\section{Conclusions\label{sec:Conclusion}}
In this paper, we derived tractable expressions for the characteristic
function of the aggregate interference power for homogeneous distribution of UEs using the \gls{SG} framework
for $N_{c}\times N_{v}$ \gls{UcylA} placed at arbitrary height. We proved that for $h>0$, the distribution of the aggregated interference could be analytically approximated by a decomposable  mixture of two distributions: skewed alpha-stable and Gaussian. The numerical analysis validates the results derived for the array configurations, including the case when $N_{v}=1$ for UCA. The analysis
of average service probability vs. \gls{UcylA} height shows that there are different trade-offs to exploit. Furthermore, the appropriate array geometry depends on different environment and propagation parameters. The impact of multipath has been evaluated analytically, thus, showing the trade-off by the increased aggregated interference, and the diversity for the UE of interest. Blockage makes the analysis to be very realistic for mmW and 6G system, and the blockage analysis has showed that there are several design insights to exploit.

Future work could consider the extension to \gls{DAS} with different multi-AP coordination. The availability of the aggregated
interference distributions in analytic form opens the possibility
to explore multi-AP cooperation that
are possible otherwise by massive simulations.

\vspace{6pt} 
\appendix
\begin{appendices}
\section{Singularity Point\label{sec:A---SingularityPoint}}
According to \cite[sections 313.14 \& 021.12]{INTEGRALTAFEL} we have $\int_{0}^{\infty}(e^{j\mu t})t^{-\alpha-1}dt=(-j\mu)^{\alpha}\Gamma(-\alpha)$ which is used in Ref.\cite{WinPintoMatThInterference} when $h=0$ for proving the alpha-stability of the distribution of aggregated interference power. However, for $h>0$  one will encounter the following integral and its solution:
\begin{equation}
    \int_{0}^{|\omega|/h^{2b}}\left[\frac{1-e^{j \mu t}}{t^{\alpha+1}}\right]dt=\lim\limits_{\varepsilon\rightarrow0}\left.(-j\mu)^{\alpha}\Gamma(-\alpha,-j\mu t)-\frac{1}{\alpha t^{\alpha}}\right]_{t=\varepsilon}^{t=|\omega|/h^{2b}}\label{eq:IncompleteGammaSolution}, 
\end{equation}
for any constant and real $\mu$ and $0<\alpha < 1$. In order to inspect the existence of any singularity point, one could use the series of incomplete Gamma function
\begin{align}
\Gamma(x,z)= & \Gamma(x)-z^{x}\sum_{k=0}^{\infty}\frac{(-z)^{k}}{(x+k)k!}.
\end{align}
Since $x=-\alpha\in\left(-1,0\right)$, there exist one singularity
point for $z\rightarrow0$ ($t\rightarrow0$ in (\ref{eq:IncompleteGammaSolution})) that is compensated by the integral $\int_{0}^{|\omega|/h^{2b}}\left[\frac{1}{t^{\alpha+1}}\right]dt$.

\section{Proof of MRC \label{sec:AppendixMRC}}
Let the combiner be
\begin{equation}
 \hat{x}  =  \sum_{\ell=1}^{L}c_{\ell}^{H}y_{\ell} = \mathbf{c}^H \, \mathbf{y},
\end{equation}
with weights $\mathbf{c}$ from a received signal $\bf{y}=\bf{h}x+{\boldsymbol{\iota}}$ where the jth entry of \textbf{h} is $h_j=\beta_j/r_j^b$ for distance $r_j$, and the \gls{CF} of the interference $I_j=|\iota_j|^2$ is known $\Psi_{\ell}(\omega)$. The MRC are designed to maximize the SIR $\Upsilon$, and thus the service probability $P_{MRC}(\mathbf{c})$ where the instantaneous SIR is
\begin{equation}
    \Upsilon = \frac{\mathbf{c}^H \, \mathbf{h \, h}^H \, \mathbf{c}}{{\mathbf{c}^H \, \mathbf{\bar{D}_{I}} \, \mathbf{c}}},\label{eq:RayleighQuotient}
\end{equation}
$\bar{D}_{I}=E_I[D_I]$ for $ D_{I} = diag(I_1,I_2,\cdots, I_L)$, and $\bar{I}_{\ell}=E[I_\ell]$ that can be derived from CF, such as for $R_{max}<\infty$. However, for skewed alpha-stable distributions for $R_{max} \to \infty$ (UCA for $h=0$, Sect.\ref{subsec:SpecificCases}), the mean does not exist, and maximization for the choice $\bar{I}_{\ell}$ as median does not change the conclusions. The Rayleigh quotient (\ref{eq:RayleighQuotient}) is known to be maximized for the choice 
\begin{equation}
 \mathbf{c}_{opt}  = \frac{\bar{\mathbf{D}}_{I}^{-1} \mathbf{h} }{\mathbf{h}^H  \bar{\mathbf{D}}_{I}^{-1} \mathbf{h}}\label{eq:QuotientEigMax}
\end{equation}
of the weights $\mathbf{c}$. The service probability reduces to
\begin{equation}
    {P}_{MRC} =\mathbb{P}_{I}  (\mathbf{c}_{opt}^{H}  {\mathbf{D}}_I \mathbf{c}_{opt} < \mathbf{c}_{opt}^{H}  \mathbf{h}  \mathbf{h}^H   \mathbf{c}_{opt} \frac{\sigma}{T}),
    \label{eq:ProbMRC}
\end{equation}
 and after some analytic it reduces to 
\begin{equation}
    {P}_{MRC}  =  \mathbb{P}_{I} \left( I_{tot} < \frac{B}{T}\right)\label{eq:MRC_Interf_Proof},
\end{equation}
where the aggregated weighted interference is $I_{tot}=\sum_{\ell=1}^{L} I_{\ell} \alpha_{\ell}$ for $\alpha_{\ell} = |\beta_{\ell}|^2/(\bar{I}_{\ell}^{2 } r_{\ell}^{2b})$ and $B = \left( \sum_{\ell=1}^{L} |\beta_{\ell}|^2/{\bar{I}_{\ell}\, r_{\ell}^{2b}}  \right)^{2}$. Thus, the analysis for the service (or complementary, for the outage) depends on the \gls{CDF} of  $I_{tot}$  and in turn on the \gls{CF}
\begin{equation}
    \Psi_{I_{tot}}(\omega)=\prod_{\ell=1}^{L}\Psi_\ell (a_{\ell}\omega)\label{eq:CFMRC}.
\end{equation}
that evidences the multiple usage in the main text for service probability analysis.

\section{Augmented Interference\label{sec:AugmentedInterference}}
By considering the geometric model used for the NLOS paths, the \gls{CF} of the aggregated interference power for an isotropic antenna (can be readily generalized for \gls{UCA} or \gls{UcylA} but it is avoided here for simplicity) by re-adapting relation (\ref{eq:CFpoint_antenna_h>0}) with small approximations, for NLOS paths calculated as $\Psi_{NLOS}(\omega)=\Psi_{1}(\omega). \Psi_{2}(\omega)$, where 
\begin{eqnarray*}
\Psi_{1}(\omega) & = & \exp\left(-\frac{\pi\lambda}{C_{\alpha}}\left|\omega\right|^{\alpha}\left|\bar{\beta}\right|^{2\alpha}P\left(-\alpha,\frac{-j|\omega|\bar{\beta}^{2}}{(h^{2} + d^{2})^{b}}\right)\left(1-j\textrm{sign}(\omega)\tan\frac{\pi\alpha}{2}\right)+\pi\lambda (h^{2} + d^{2})\right), \\
\Psi_{2}(\omega) & = & \exp\left(+\frac{\pi\lambda \, d}{C_{\alpha}}\left|\omega\right|^{\frac{\alpha}{2}}\left|\bar{\beta}\right|^{2\alpha}P\left(\frac{-\alpha}{2},\frac{-j|\omega|\bar{\beta}^{2}}{(h^{2} + d^{2})^{b}}\right)\left(1-j\textrm{sign}(\omega)\tan\frac{\pi\alpha}{4}\right)+2\pi\lambda \,d \sqrt{h^{2} + d^{2}}\right). 
\end{eqnarray*}
Now, the CF of the total augmented aggregate interference is
\begin{equation}
    \Psi_{{{I}}}(\omega) = \Psi_{NLOS}^{(L-1)}(\omega)\Psi_{LOS}(\omega)
\end{equation}
where $\Psi_{LOS}(\omega)$ is as derived as in relation (\ref{eq:CFpoint_antenna_h>0}) and $L$ is the total number of paths. The CDF of the augmented interference can be achieved from this CF.
\end{appendices}
\bibliographystyle{IEEEtran}
\bibliography{DAS_Ref}
\pagebreak{}

\end{document}